\documentclass[journal,10pt]{IEEEtran}
\usepackage{graphicx}
\usepackage{float}
\usepackage{stfloats}
\usepackage{subcaption}
\usepackage{amsmath}
\usepackage{amsfonts}
\usepackage{pdfpages}
  \usepackage{caption}
  \usepackage{enumitem}
  \usepackage{multirow}
  \newcommand{\subscript}[2]{$#1 _ #2$}
\usepackage{algpseudocode}
\usepackage{multicol}
\usepackage{cite}
\usepackage[ruled,linesnumbered]{algorithm2e}
\ifCLASSINFOpdf
\else
\fi
\begin{document}
\title{A Bayesian Poisson-Gaussian Process Model for Popularity Learning in Edge-Caching Networks}
\author{Sajad~Mehrizi,
	Anestis~Tsakmalis,~\IEEEmembership{Student~Member,~IEEE,}
	Symeon~Chatzinotas,~\IEEEmembership{Senior~Member,~IEEE,}
	and~Bj{\"o}rn~Ottersten,~\IEEEmembership{Fellow,~IEEE}
\thanks{All authors are with Interdisciplinary Centre for Security, Reliability and Trust (SnT), University of Luxembourg}}
%	{$\left\{ sajad.mehrizi, anestis.tsakmalis, symeon.chatzinotas, bjorn.ottersten \right\}$$@uni.lu$ }}% <-thi
\maketitle
%%%%%%%%%%%%%%%%%%%%%%%%%%%%%%%%%%%%%%%%%%%%%%%%%%%%%%%%%%%%%%%%%%%%%%%%%%%%%%%
\begin{abstract}
Edge-caching is recognized as an efficient technique for future cellular networks to improve  network capacity and  user-perceived quality of experience. 
To enhance the performance of caching systems, designing an accurate content request  prediction algorithm plays an important role. In this paper, we  develop a flexible model, a Poisson regressor based on a Gaussian process,   for the content request distribution. 
  The first important advantage of the proposed model is that it encourages  the already existing or \textit{seen}  contents with similar features to be correlated in the feature space and therefore it acts as a regularizer for the estimation. Second, it allows to predict the popularities of newly-added or \textit{unseen} contents whose statistical data is not available in advance. In order to learn the model parameters, which yield the Poisson arrival rates or alternatively the content \textit{popularities}, we invoke the Bayesian approach which is robust against over-fitting.
 However, the resulting posterior distribution is  analytically intractable to compute. To tackle this, we apply a Markov Chain Monte Carlo (MCMC) method to  approximate this distribution which is also asymptotically exact. Nevertheless, the MCMC is computationally demanding especially when the number of contents is large. Thus, we employ the Variational Bayes (VB) method as an alternative low complexity solution. More specifically, the VB method addresses the approximation of the posterior distribution through an optimization problem. Subsequently, we present a fast block-coordinate descent algorithm to solve this optimization problem. Finally, extensive simulation
 results both on synthetic and real-world datasets are provided to show the accuracy of our prediction algorithm and the cache hit ratio (CHR) gain compared to existing methods from the literature.
\end{abstract}
%%%%%%%%%%%%%%%%%%%%%%%%%%%%%%%%%%%%%%%%%%%%%%%%%%%%%%%%%%%%%%%%%%%%%%%%%%%%%%%%%%%
%\IEEEpeerreviewmaketitle
\section{Introduction}
Recently, there has been a tremendous growth in mobile data traffic due to the increasing number of mobile devices and growing user interest towards bandwidth-hungry applications (e.g., 4K videos). According to a recent Cisco report~\cite{indexglobal}, it is predicted that from 2016 to 2021 mobile traffic will increase at a $47 \%$ compound annual growth rate (CAGR), two times faster than the growth of global IP fixed traffic during the same period. Nevertheless, such an inevitable growth has raised concerns about the flow of wireless traffic that traditional network architectures can tolerate. To cope with this, much effort has have been devoted towards designing and developing the 5th generation (5G) wireless cellular systems. The 5G system must provide fast, flexible, reliable and continuous wireless connectivity, while supporting the growing mobile traffic. 
  
In this respect, caching popular contents at the network edge has been introduced ~\cite{shanmugam2013femtocaching, ge20145g} as an efficient solution to tackle the aforementioned issues. By observing that  only a small number of popular contents are
frequently requested by users, bringing these contents  from the core network closer to the end mobile users avoids  downloading the same content multiple times through the backhaul links. As a result, by serving users locally, edge-caching can jointly  increase connectivity,  reduce
the delay,  alleviate the backhaul link congestion and  improve quality of service (QoS) of mobile users.
  
 %here you can highlight that the majority of early work was focused on optimizing placement and delivery taking into account that popularities are perfectly known

There has been a growing research interest in edge-caching networks,  the majority of which has  focused on the development and the performance analysis of various cache placement and delivery strategies taking into account that popularities are perfectly known. For example, in~\cite{shanmugam2013femtocaching} a cache placement algorithm has been proposed to minimize the expected downloading time for contents. In a multiple base station scenario, the authors of~\cite{chen2017cooperative} studied a probabilistic caching policy to increase content
diversity in the caches.  In~\cite{peng2014joint}, physical layer features are used in the cache placement problem to minimize network cost while satisfying users' QoS requirements. The authors in~\cite{vu2018edge} investigated energy efficiency and delivery time of an edge-caching network. In addition, various coding schemes, intra and inter sessions, have been proposed to enhance caching performance~\cite{shanmugam2013femtocaching,maddah2014fundamental,han2016phy}. 

Still, the performance of caching schemes in the first place depends on popularity estimation accuracy. Therefore, understanding content popularity  is of great importance. In real life, there are various factors that influence the way users consume contents such as social interactions, culture, user profiles, content features and so on. Nevertheless, all the underlying factors that affect users to consume contents might be either difficult to model or unavailable at the edge network (e.g. due to privacy issues). Hence, discovering the hidden content request pattern is a challenging task.    Another important issue is that content producers constantly introduce new contents to the system. Proactively caching these contents can benefit both users and content providers. However, statistical data for these new or unseen contents is rarely available which makes proactive caching difficult.  

\subsection{Related Work}
During the recent years, several papers investigated machine learning techniques to predict the content popularity.  In this context, the popularity learning problem can be categorized in two general approaches: model-free and model-based. In the model-free approach, there is no assumption on the content request distribution. The popularity learning is then performed within the process of optimizing a reward function (e.g cache hit ratio) by the so-called exploration-exploitation procedure. Multi-armed-bandit (MAB) and reinforcement learning algorithms are mostly based on this approach which also have been adapted to  edge-caching applications~\cite{muller2017context,song2017learning,sadeghi2018optimal,somuyiwa2018reinforcement}. On the other hand, in the model-based approach, it is assumed that the content requests are generated by a parametric distribution. The Poisson stochastic process is a popular model adopted in content delivery networks~\cite{garetto2016unified} and has also been used in edge-caching~\cite{bharath2016learning,elbamby2014content}. Once the request generation process is modeled, the next step is to estimate the popularity. A simple way is to take the average of the instantaneous requests, which is equivalent to the maximum likelihood estimation (MLE) from the estimation theory perspective. MLE performs well when the size of request samples is large. %For example, the authors in~\cite{bacstuug2015big} assumed that a large amount of data is available for popularity estimation. 
However, as it is reported
in~\cite{paschos2016wireless}, a  base station cache typically may receive 0.1 requests/content/day which is too small in comparison with a typical content delivery network cache which normally receives 50 requests/content/day. This indicates that MLE provides inaccurate estimation of content popularity in the local caches. 

To improve the popularity estimation accuracy, side information  (user profile and content features) can also be incorporated in learning algorithms. In~\cite{bacstuug2015transfer,bharath2016learning}, users' social interactions are leveraged to speed up the learning convergence rate. One important issue with this kind of side information is that users may not be willing to share their personal profiles with the entity operating the edge caches. On the other hand, content features (e.g topic categories) can  easily and cheaply be obtained from the content server without jeopardizing users' privacy. In~\cite{doan2018content}, the authors used the content features to predict the popularity of unseen contents.  More recently, the content features are used to  learn the user-level preferences~\cite{jiang2018user}.  However, due to the data scarcity issue in the local caches, the estimation is prone to overfitting and may not be accurate. Most importantly though, it is widely admitted that in order to have  accurate prediction, a more complex model is needed. This will help the network operator to identify underlying hidden structures of the content request generative process, discover popular but unseen contents and disseminate them optimally in order to improve the content delivery in the network.
 
\subsection{Contributions}
 In this paper, we take content features  into account and introduce a new sophisticated probabilistic model for the content requests. The learning process is performed in the Bayesian paradigm which is robust against overfitting and provides a way to quantify our uncertainty about the estimation. The model allows us to define  different types of predictive distributions by which we can effectively model the uncertainty of future requests. The statistical information of these posterior predictive distributions can potentially be  used to enhance the performance of a caching policy. %Here, we should also mention that the central contribution of this paper is not to devise  a caching policy but rather  to propose a more accurate   probabilistic model for content requests.
Overall, the main contributions of the paper are summarized as:
 
 \begin{itemize}
 	\item We provide a probabilistic model for stationary content requests which exploits the similarities between contents.  This model can perform two important tasks. First, it encourages the seen contents with similar features to have close popularities. In other words, it acts as a regularizer and as a result it provides a better estimation accuracy. Therefore, our model is much more flexible for popularity prediction than the method proposed in~\cite{doan2018content} where features are used only to predict the popularity of  unseen content. Second, similar to~\cite{doan2018content} but in a more efficient way,  our model can predict the popularity of unseen contents where statistical information is not available in advance.

 	%We provide a probabilistic model, a Poisson regressor based on a Gaussian process, for stationary content requests which employs the similarity between contents. The Gaussian process is  a very flexible and powerful statistical structure that can model complex nonlinear relationships between the popularities and the features. This model can perform two important tasks. First, it encourages the seen contents with similar features to have close popularities. In other words, it acts as a regularizer and as a result it provides a better estimation accuracy. Therefore, our model is much more flexible for popularity prediction than the method proposed in~\cite{doan2018content} where features are used only to predict the popularity of a an unseen content. Second, similar to ~\cite{doan2018content} but in a more efficient way,  our model can predict the popularity of unseen contents whose statistical information is not available in advance.

 	\item  We introduce a Poisson regressor based on a Gaussian process to  model  the content requests. The Gaussian process is  a very flexible nonparametric statistical structure that can model complex nonlinear relationships between the popularities and the features. In addition,  a Bayesian method is developed to learn the parameters of the proposed probabilistic model. Due to few content request samples in the local cache, Bayesian learning provides a powerful framework to mitigate overfitting.
 	\item Since there is no closed form solution for the posterior distribution, two inference algorithms are presented. First, the HMC method is used to estimate the posterior distribution which  asymptotically provides an exact solution. Because the HMC can be computationally demanding, we introduce the VB method which turns the inference problem into an optimization. To efficiently solve this, an algorithm based on a combination of  coordinate descent  and parallel computing is developed.% In addition, due to the model structure, both methods can be easily implemented for on-line learning i.e. the content requests arrive sequentially over time.
 \end{itemize}

This paper is organized as follows. The system model  and problem statement are described in Section \ref{seq:Sysmodel}. In Section \ref{sec:conMoledl},  our probabilistic model is introduced. In Section \ref{BayesinInfer}, we apply  the  Bayesian approach where two methods, namely the HMC and the VB, are presented  for the inference task. Finally, Section \ref{SimRes} shows the simulation results and Section \ref{Conclusion} concludes the paper.

\textit{Notation}: lower- (upper-) case boldface letters denote column vectors (matrices), whose ($i$,$j$)-th entry is represented by ${\left[ . \right]_{ij}}$. ${\left( . \right)^{ - 1}}$ and ${\left( . \right)^{T}}$ denote inverse and transpose, respectively.

\begin{figure}
	\centering
	\includegraphics[scale=0.4, trim=0 0 0 0]{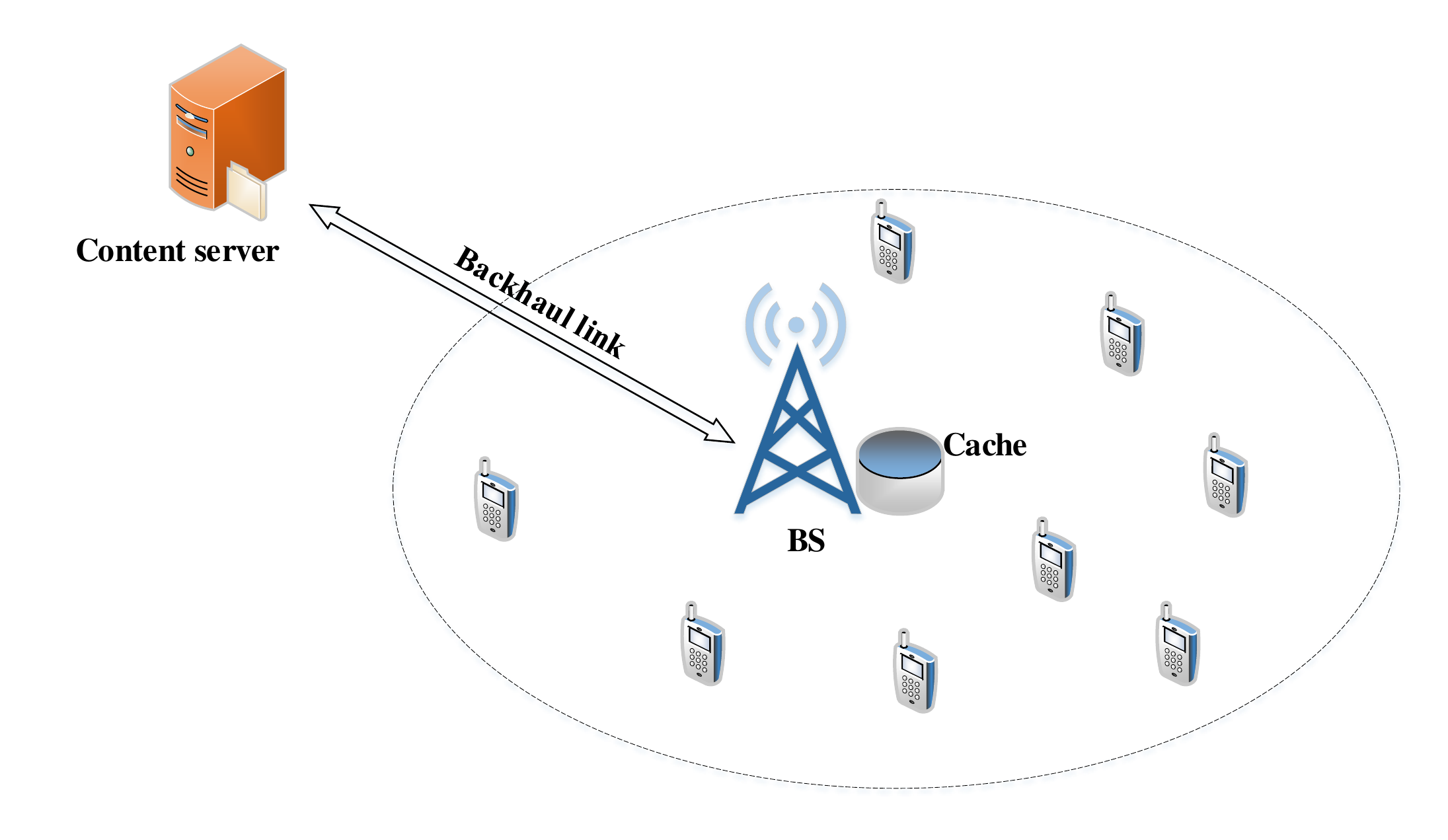}%SysModel
	\caption{The system model of a cache-enabled  cellular network}
	\label{fig:SysModel}
\end{figure}
%%%%%%%%%%%%%%%%%%%%%%%%%%%%%%%%%%%%%%%%%%%%%%%%%%%%
%%%%%%%%%%%%%%%%%%%%%%%%%%%%%%%%%%%%%%%%%%%%%%%%%%%%
\section{System Model and problem formulation}\label{seq:Sysmodel}
In this paper, we consider a cache-enabled  cellular network consisting of a base station (BS) which serves  mobile users within its coverage area as shown in Fig.\ref{fig:SysModel}. There are $U$ mobile users in the cell, where each user makes random requests from
a library of contents ${\cal C} = \left\{ {{c_1},...,{c_M}} \right\}$, where $M$ is the total number of contents. 

The BS is equipped with a cache which can store a certain number of contents depending on
its storage capacity. Moreover, the BS is connected to a remote content server which has access to the whole content library ${\cal C}$ through the backhaul links. Time is divided into different time slots and at each time slot\footnote{The time slots can be hours, days, etc.}, each user independently requests a content (or contents)\footnote{There is no limitation on the number of requests by a user at a time slot. Even, in practice, a user may request the same contents more than once, for example, he/she may  watch a Youtube video multiple times during a day, assuming that a time slot is one day.} from the library ${\cal C}$. To alleviate the traffic burden on the backhaul links and increase the users' QoS, some contents are stored in the cache depending on the caching policy. The requested contents by the users will be served directly if they are already cached; otherwise they are fetched from the content server. We suppose that the cache module of the BS can only monitor the number of user requests towards contents of the library and cannot or is not allowed to perform any user profiling. In addition, it is assumed that the content popularity is fixed (we can assume it does not considerably change over short time intervals, e.g. a few days or weeks) and the requests are samples generated from a stationary distribution.
%%%%%%%%%%%%%%%%%%%%%%%%%%%%%%

%%%%%%%%%%%%%%%%%%%%%%%%%%%%%%%%%
We define ${{\bf{d}}_c}\left[ {{T_n}} \right] = \left[ {{d_{{c_1}}}\left[ {{T_n}} \right],...,{d_{{c_M}}}\left[ {{T_n}} \right]} \right]^T$ to be the request vector where $d_{{c_m}}\left[ {{T_n}} \right]$ is the total number of requests for content $m$ during time slot $n$ with duration $T_n$. For simplicity, we assume that $T_n = T_{n'}$ $\forall n' \ne n$ . Therefore, we can drop $T$ and show the request vector by ${{\bf{d}}_{c,n}} = \left[ {{d_{{c_1},n}},...,{d_{{c_M},n}}} \right]^T$.  Also,  the requests for $n' \ne n$   are  presumed to be statistically independent random variables. A common parametric model for  the requests  is  the Poison stochastic process i.e. ${{\bf{d}}_{c,n}}\sim Poi\left( {\bf{r}} \right),\forall n = 1,2,...$, where ${\bf{r}}$ is the Poisson arrival rate or the content popularity. Here, we should also mention that the Zipf distribution, which is widely used in the  literature (e.g.~\cite{breslau1999web}), is not a distribution to model the  requests. Specifically, it only models the  \textit{ordered means} of requests not the \textit{requests} themselves. 

 Any caching algorithm needs an estimate of content popularities  ${\bf{r}} = E\left\{ {{{\bf{d}}_c}} \right\}$ to operate. For example, a common caching strategy is to maximize the average CHR:
\begin{subequations}\label{eq:cachePolicy}
\begin{align}
&\mathop {\max }\limits_{\bf{w}} \,{{\bf{w}}^T}{\bf r}\\
&s.t:{{\bf{w}}^T}{\bf{s}} \le C\\
&{\bf{w}} \in {\left\{ {0,1} \right\}^{M \times 1}}\label{BinCons}
\end{align}
\end{subequations}
where  $C$ is the cache size and $\bf w$ is the cache design variable. %The problem is combinatorial and NP hard to solve. However, by relaxing the binary constraint \eqref{BinCons} to the continuous bounded set ${{\rm{[0,1]}}^{M \times 1}}$ problem \eqref{eq:cachePolicy} turns into a linear program which is easy to solve.

 A simple way to approximate  the popularity of a content is  the average number of requests within $N$ observation time slots: 
\begin{equation}\label{eq:MeanPoisson}
{r_m} \approx \frac{{\sum\limits_{n = 1}^N {{d_{{c_m},n}}} }}{N},\quad \forall m = 1,...,M
\end{equation}
%where  $N$ is the total number of  request observations during the training period.
 %Hence, the popularity of a content is defined as the expected number of requests within time interval $T_n$ which is computed as: 
%\begin{equation}\label{eq:MeanPoisson}
%{r_m} = \frac{{\sum\limits_{n = 1}^N {{d_{{c_m},n}}} }}{N},\quad \forall m = 1,...,M
%\end{equation}
%where ${r_m}$ is the popularity of content $c_m$ and $N$ is the total number of  request observations during the training period. Here, we should also mention that the Zipf distribution, which is well known in the  literature, is not a distribution to model the  requests. Specifically, it only models the  \textit{ordered means} of requests not the \textit{requests} themselves.

Eq. \eqref{eq:MeanPoisson} is equivalent to the MLE  approach for popularity estimation which is very simple, but it has some flaws. First, it suffers from severe overfitting especially when the training set has only a few request observations. Second, it cannot incorporate any kind of side information. For example, users commonly request contents based on their features and therefore we expect content popularities to be correlated in their feature space. By appropriately using this underlying prior knowledge about requests, popularity estimation precision can be significantly improved. Learning the correlation in the content feature space can also help to predict the popularity of an unseen content. 

%In practice,  there are many factors that influence the interest of a user toward specific contents. For example, users belong to the same aging group usually are interested in  same kinds of movies genre.  In addition, noways, users are typically triggered by different types of social  networks such as Facebook, Tweeter and so on which make many users to follow the same request pattern.  Users in a specific region may requests the same types of contents meaning that there is spatial correlation between requests. For example, in research areas that there are many universities and researcher centers users will request contents related to academic topic.  In summary there is a very complex structure behind the popularities of contents which is affected by content topics, users profiles, social belonging and also users spatial information. However, due to concern about their privacy the cache entity doesn't have access to information form users side and only can have content features. 
%%%%%%%%%%%%%%%%%%%%%%%%%%%%%%%%%%%%%%
\begin{figure}
	\centering
	\includegraphics[scale=0.65, trim=0 0 0 0]{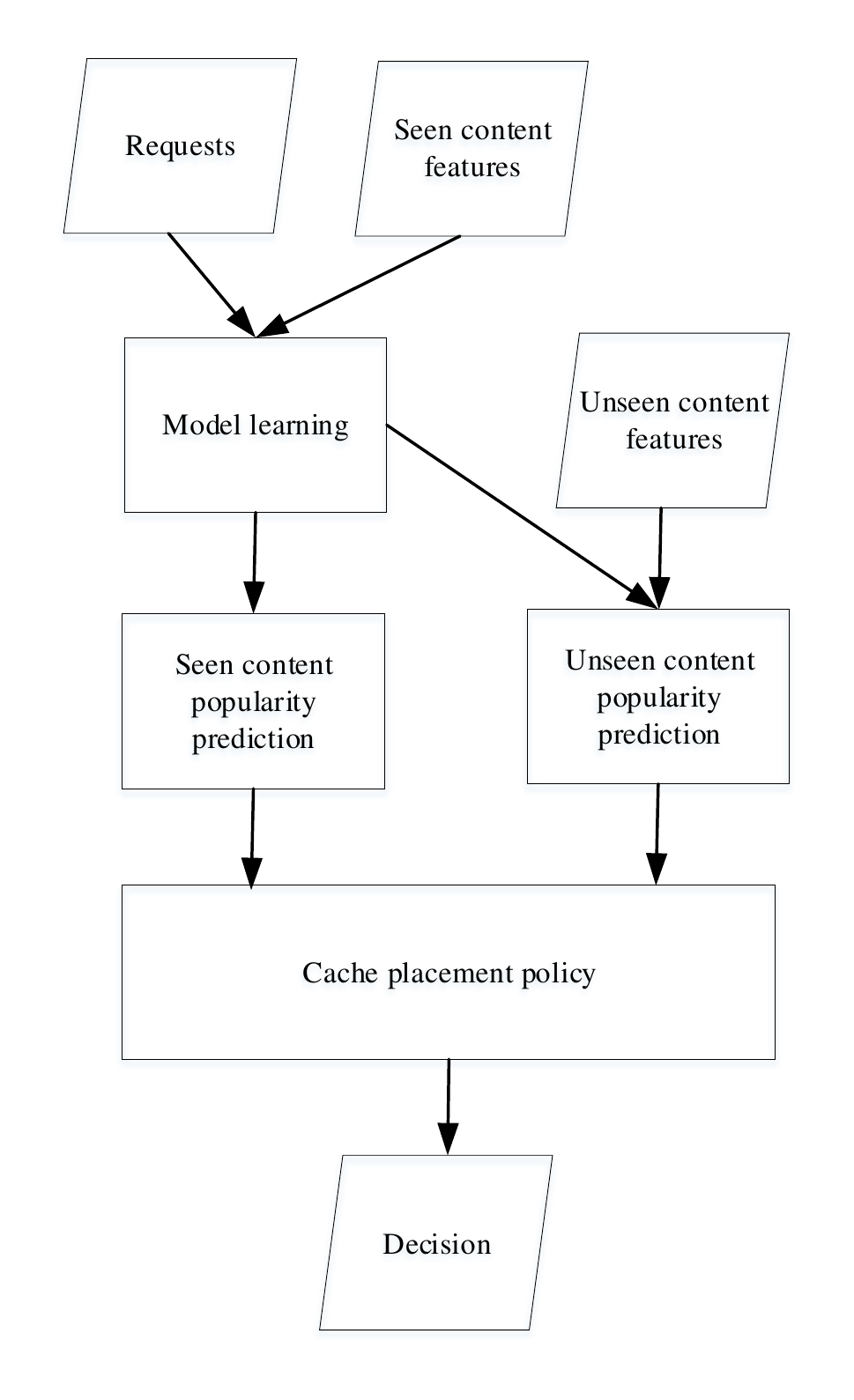}%placement_learning
	\caption{Illustration of content popularity learning and caching }
	\label{fig:flowchartLearningCaching}
\end{figure}
%%%%%%%%%%%%%%%%%%%%%%%%%%%%%%%%%%%%%%%
To overcome these issues, we adopt the Bayesian modeling scheme.  In this approach, both data, $\bf d$, and parameters, $\bf r$, are considered to be random variables and the first step   is to define a joint distribution over them: 
\begin{equation}
p\left( {{\bf{d}},{\bf{r}}} \right) = \underbrace {p\left( {{\bf{d}}|{\bf{r}}} \right)}_{\begin{array}{*{20}{c}}
	{{\rm{data}}\,\,\,\,\,{\rm{generative}}}\\
	{{\mkern 1mu} {\kern 1pt} {\mkern 1mu} {\kern 1pt} {\mkern 1mu} {\kern 1pt} {\mkern 1mu} {\kern 1pt} {\mkern 1mu} {\kern 1pt} {\mkern 1mu} {\kern 1pt} {\mkern 1mu} {\kern 1pt} {\mkern 1mu} {\kern 1pt} {\mkern 1mu} {\kern 1pt} {\mkern 1mu} {\kern 1pt} {\mkern 1mu} {\kern 1pt} {\rm{pdf}}}
	\end{array}}\underbrace {p\left( {{\bf{r}};{\boldsymbol{\zeta }}} \right)}_{\begin{array}{*{20}{c}}
	{{\rm{prior}}}\\
	{{\mkern 1mu} {\kern 1pt} {\rm{pdf}}}
	\end{array}}
\end{equation}
where the data generative distribution models the way data is generated given the parameters and the prior distribution represents the initial uncertainty of the parameters. The prior distribution   may also have some  parameters, $\boldsymbol{\zeta }$, which are called the hyper-parameters and usually assumed  deterministic.

Fig. \ref{fig:flowchartLearningCaching} presents our work-flow scheme for popularity learning and caching. In the model learning block, the proposed probabilistic model is trained based on the requests and the features of seen contents. In the popularity predicting blocks, the popularities of both seen and unseen content are predicted. Finally, a decision about which content should be cached is taken in the cache placement policy block.
\section{The content request model}\label{sec:conMoledl}
 In this section, we present our  Bayesian probabilistic model. Before introducing the model, we summarize the basic concepts of Gaussian processes which are essential for the subsequent sections.
\subsection{Gaussian Processes in a Nutshell}\label{seq:GP}
Gaussian processes are  powerful non-parametric Bayesian tools suitable for modeling real-world problems.
A Gaussian process is a collection of random variables, any finite number of which have a joint Gaussian distribution. Using a Gaussian process, we can define a distribution over non-parametric functions $f\left( {\bf{x}} \right)$:
%%%%%%%%%%%%%%%%%%%%%%%%%%%%%%%%%%%%
\begin{equation}\label{GPdef}
f\left( {\bf{x}} \right)\sim {\cal G}{\cal P}\left( {{{\mu }}\left( {\bf{x}} \right),{{K}}\left( {{\bf{x}},{\bf{x'}}} \right)} \right)
 \end{equation}
 %%%%%%%%%%%%%%%%%%%%%%%%%%%%%%%%%%%%
where $\bf x$ is an arbitrary  input variable with $Q$ dimensions, and the  mean function, ${{{\mu }}\left( {\bf{x}} \right)}$, and the Kernel function, $K\left( {{\bf{x}},{\bf{x'}}} \right)$,   are respectively defined as:
\begin{align}
&{\bf{\mu }}\left( {\bf{x}} \right) = E\left[ {f\left( {\bf{x}} \right)} \right]\\
&K\left( {{\bf{x}},{\bf{x'}}} \right) = E\left[ {\left( {f\left( {\bf{x}} \right) - \mu \left( {\bf{x}} \right)} \right)\left( {f\left( {{\bf{x'}}} \right) - \mu \left( {{\bf{x'}}} \right)} \right)} \right].
\end{align}
This means that a collection of $M$ function value samples has a joint Gaussian distribution:
\begin{equation}
\left[ {f\left( {{{\bf{x}}_1}} \right),...,f\left( {{{\bf{x}}_M}} \right)} \right]^T\sim {\cal N}\left( {{\boldsymbol{\mu }},{\bf{K}}} \right)
\end{equation}
where ${\boldsymbol{\mu }} = {\left[ {\mu \left( {{{\bf{x}}_1}} \right),...,\mu \left( {{{\bf{x}}_M}} \right)} \right]^T}$ and the covariance matrix ${\bf K}$ has entries ${\left[ {\bf{K}} \right]_{i,j}} = K\left( {{{\bf{x}}_i},{{\bf{x}}_j}} \right)$.
The kernel function specifies the main characteristics of the function that we wish to model and the basic assumption is that variables $\bf x$ which are close are likely to be correlated. Constructing a good kernel function  for a learning task depends on intuition and experience. In this paper, we only focus on a popular and simple kernel which is the squared exponential kernel (SEK). However, our methodology can easily be applied to other types of Kernels. The SEK has the following form:
 \begin{equation}\label{eq:SEK}
K\left( {{{\bf{x}}_i},{{\bf{x}}_j}} \right) = {\alpha _0}{e^{ - \sum\limits_{q = 1}^Q {{\alpha _q}{{\left\| {{x_{q,i}} - {x_{q,j}}} \right\|}^2}} }}
 \end{equation}
 where ${\alpha _0}$ is the vertical scale variation and ${\alpha _{q }}$ is the horizontal scale variation on dimension $q$  of the function. By using  different scales for each input dimension, we vary their importance. If ${\alpha _{q }}$ is close to zero, dimension $q$ will have little influence on the  covariance of function values. 
 The covariance function \eqref{eq:SEK} is infinitely differentiable and is thus very smooth. More details about Gaussian processes and kernel functions can be found in~\cite{rasmussen2004gaussian}.
 %%%%%%%%%%%%%%%%%%%%%%%%%%%%%%%%%%%%%%%%%%%%%%%%%%%%%%%
 \begin{figure}
 	\centering
 	\includegraphics[scale=0.7, trim=0 0 0 0]{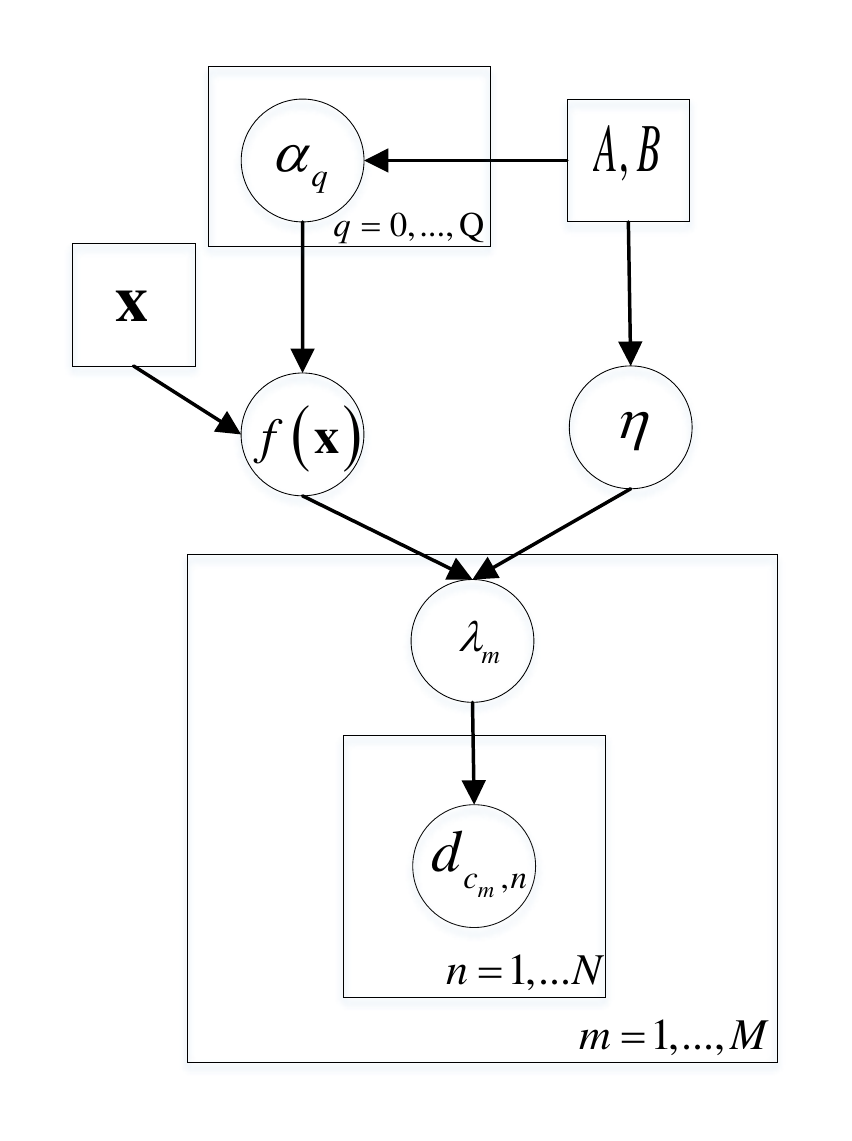}%,DrawingModel
 	\caption{The proposed probabilistic model for content requests}
 	\label{fig:probmodel}
 \end{figure}
%%%%%%%%%%%%%%%%%%%%%%%%%%%%%%%%%%%%%%%
 \subsection{The proposed probabilistic model}\label{seq:propModel}
 In this subsection, we introduce our probabilistic model for content requests.
 Each content is  assumed to have a set of features. For instance, YouTube videos may belong to specific categories (e.g education, art, entertainment, science-technology,.. )  and have some other features such as release year, language and so on. We let ${{\bf{x}}_m}$ be the feature vector of content $c_m$ with $Q$ dimensions whose values can be either binary or continuous. 
 The proposed regression-based hierarchical (multilevel) probabilistic model is the following:
 \begin{subequations}\label{eq:CellGPdef}
 	\begin{align}
 	&{d_{{c_m},n}}|{\lambda _m}\left( {{{\bf{x}}_m}} \right)\sim Poi\left( {{e^{{\lambda _m}\left( {{{\bf{x}}_m}} \right)}}} \right),\forall n = 1,...,N \label{eq:Genlev1}\\
 	&{\lambda _m}\left( {{{\bf{x}}_m}} \right)|f\left( {{{\bf{x}}_m}} \right),{\eta }\sim {\cal N}\left( {f\left( {{{\bf{x}}_m}} \right),{\eta }} \right)\label{eq:Genlev2}\\
 	&f\left( {\bf{x}} \right)|{\bf{x}},{\alpha _0},...,{\alpha _{Q}}\sim {\cal GP}\left( {0,K\left( {{\bf{x}},{\bf{x'}}} \right)} \right) \label{eq:Genlev3}\\
 	&\eta ,{\alpha _0},...,{\alpha _Q}\sim Gam\left( {A,B} \right).\label{eq:Genlev4}
 	\end{align}
 \end{subequations}

 The first level of the model, \eqref{eq:Genlev1},  is the observation   distribution for content requests.  At this level, the request for content $c_m$   is assumed to follow a Poisson distribution with  natural parameter ${\lambda _m}\left( {{{\bf{x}}_m}} \right)$ which  is a function of its features. We note here that the request rate is an exponential function of the natural parameter, ${r_m}\left( {{{\bf{x}}_m}} \right) = {e^{{\lambda _m}\left( {{{\bf{x}}_m}} \right)}}$. As we previously mentioned, it is expected that there is a similar request pattern  between contents with similar features. This prior information is employed at the higher levels.
 In \eqref{eq:Genlev2}, ${\lambda _m}\left( {{{\bf{x}}_m}} \right)$ follows a normal distribution with mean $f\left( {{{\bf{x}}_m}} \right)$ and variance ${\eta }$. By this assumption, we allow contents with exactly the same features to have different popularities which is possible in practice. At the higher level of the model, \eqref{eq:Genlev3}, we assume that $\left\{ {f\left( {{{\bf{x}}_m}} \right)} \right\}_{m = 1}^M$ are realizations of function $f\left( {\bf{x}} \right)$ drawn from a Gaussian process with zero mean and kernel function $ K$. By this assumption, contents with similar features are encouraged to be correlated in the feature space.  Finally, in level \eqref{eq:Genlev4}, we introduce   uncertainty in the values of ${\boldsymbol{ {{\theta }} }} = {[\eta ,{\alpha _0},...,{\alpha _Q}]^T}$ since in practice the available prior knowledge may not be enough to fix them. Therefore, we use a Gamma distribution, $Gam\left( {A,B} \right)$, as a prior for them where $A$ and $B$  are respectively the shape and scale   parameters.  The corresponding graphical model of \eqref{eq:CellGPdef}  is depicted in Fig. \ref{fig:probmodel}.  The plates represent multiple samples of random variables. The unshaded circle nodes indicate unknown quantities and the squares show the deterministic parameters of the model.
 
 We should mention that the line separating the priors and the generative distribution in model \eqref{eq:CellGPdef} depends on the question in hand. More specifically, the generative distribution for content requests is Poisson in \eqref{eq:Genlev1} while the other distributions consist the prior for the natural parameter of the Poisson. On the other hand, the generative distribution for the popularities is a Gaussian process in \eqref{eq:Genlev2} and \eqref{eq:Genlev3}, while \eqref{eq:Genlev4} is the prior for the parameters of the Gaussian process. This separation lets us define two predictive distributions: for the seen contents and the unseen ones. This part is further explained in Section \ref{sec:prediction}.

 The proposed model is very flexible and can be easily extended. For example, it can be developed to model the content requests on user-level. Assume that for  user $ u$ in the cell there is vector ${\bf p}_u$ with dimensions $P$ which contains his profile information. The following probabilistic model can be used as a generative distribution for one user's requests:
 \begin{subequations}\label{eq:GPuserLevel}
 	\begin{align}
 &{d_{{c_m},u,n}}|{\lambda _{m,u}}\left( {{{\bf{x}}_m},{{\bf{p}}_u}} \right)\!\sim\! Poi( {{e^{{\lambda _m}\left( {{{\bf{x}}_m},{{\bf{p}}_u}} \right)}}} ),\forall n = 1,...,N\\
 &{\lambda _m}\left( {{{\bf{x}}_m},{{\bf{p}}_u}} \right)|f\left( {{{\bf{x}}_m},{{\bf{p}}_u}} \right),{\eta _{u}}\sim {\cal N}\left( {f\left( {{{\bf{x}}_m},{{\bf{p}}_u}} \right),{\eta _{u}}} \right)\\
 &f\left( {{{\bf{x}}},{{\bf{p}}}} \right)|{\bf{x}},{\bf{p}}\sim {\cal{GP}}\left( {0,K\left( {{{\bf{x}}},{{\bf{p}}},{{\bf{x}'}},{{\bf{p}'}}} \right)} \right).
 \end{align}
 \end{subequations}
 with the new kernel function defined as:
 \begin{align}\label{eq:userKernel}
&K\left( {{{\bf{x}}_m},{{\bf{p}}_u},{{\bf{x}}_{m'}},{{\bf{p}}_{u'}}} \right) =\nonumber\\
& {\alpha _{0,u}}{e^{ - \sum\limits_{p = 1}^P {{\beta _{pu}}{{\left\| {{p_{p,u}} - {p_{p,u'}}} \right\|}^2}}  - \sum\limits_{q = 1}^Q {{\alpha _{qu}}{{\left\| {{x_{q,m}} - {x_{q,m'}}} \right\|}^2}} }}.
 \end{align}
 Similarly to \eqref{eq:CellGPdef}, we can use Gamma distributions to model the uncertainty in the parameters of the kernel function.  The model in \eqref{eq:GPuserLevel} allows the  popularities to be correlated in the joint space of user and content features. Because, in practice, users may have different tastes (like or dislike) about the content features, we let  the kernel parameters  be different for each user in the content feature space. % In other words, users with similar profiles will request the same types of contents. 

However, there is a major issue that may discourage us to use model~\eqref{eq:GPuserLevel}. Due to privacy concerns, user profiles may not be available at the BS and therefore from now on we only focus on the cell-level content request model \eqref{eq:CellGPdef}. In the next section, we show how to learn this model and make predictions based on it.
%%%%%%%%%%%%%%%%%%%%%%%%%%%%%%%%%%%%%%%%%%%%%%%%%%%%%%%%%%%
\section{Model Learning}\label{BayesinInfer}
In this section, we utilize the Bayesian framework to learn the probabilistic model in \eqref{eq:CellGPdef}. In other words, given the content request observations ${\cal D} = \left\{ {{{\bf{d}}_{c,n}}} \right\}_{n = 1}^N$, we aim to update our uncertainty  about the  model's parameters, $\left\{ {{\lambda _m}\left( {{{\bf{x}}_m}} \right)} \right\}_{m = 1}^M,f\left( {\bf{x}} \right),{{\boldsymbol{ \theta }}}$. However, we cannot estimate the infinite-dimensional function $f\left( {\bf{x}} \right)$ and hence the focus is only on the realizations  $\left\{ {f\left( {{{\bf{x}}_m}} \right)} \right\}_{m = 1}^M$.
Moreover, to simplify the inference, we  can integrate out  ${f\left( {{{\bf{x}}_m}} \right)}$ from the model. By doing this, we have: 
\begin{equation}
{\boldsymbol{\lambda }} = {\left[ {{\lambda _1}\left( {{{\bf{x}}_1}} \right),....,{\lambda _M}\left( {{{\bf{x}}_M}} \right)} \right]^T}\sim {\cal N}\left( {{\bf{0}},{\bf{\tilde K}}} \right)
\end{equation}
where ${ \bf{\tilde K}} = {\bf{K}} + {\eta}{\bf{I}}$.
%%%%%%%%%%%%%%%%%%%%%%%%%%%%%

The inference of all unknown parameters of the  model is given by the Bayes rule as:
\begin{align}\label{eq:BayesRule}
p\left( {{\boldsymbol{\lambda }},{\boldsymbol{\theta }}|{\cal D}} \right) = \frac{{p\left( {{\cal D}|{\boldsymbol{\lambda }}} \right)\,p\left( {{\boldsymbol{\lambda }}|{\boldsymbol{\theta }}} \right)\prod\limits_{q = 0}^{Q + 1} {p\left( {{\theta _q}} \right)} }}{Z}
\end{align}
where $p\left( {{\cal D}|{\boldsymbol{\lambda }}} \right) = \prod\limits_{n = 1}^N {\prod\limits_{m = 1}^M {p\left( {{d_{{c_m},n}}|{\lambda _m}} \right)} } $ , $p\left( {{\boldsymbol{\lambda }},{{\boldsymbol{ \theta }}}|{\cal D}} \right)$ is the posterior distribution and
 the denominator $Z$ is a normalization constant also called the marginal likelihood. 
 Nevertheless, the normalization constant is intractable to compute and additionally there is no closed-form expression for the posterior distribution. So, instead, we use a Markov Chain Monte Carlo (MCMC) method to approximate the posterior distribution.  The goal of MCMC methods is to generate a set of independent samples from the target posterior distribution with enough samples to perform accurate inferences. Specifically, here, we use the Hamiltonian Monte Carlo (HMC) method which has been one of the most successful HMC methods to sample from an unnormalized distribution. Next, we give a brief overview of the HMC whose complete description can be found in~\cite{neal2011mcmc}.

HMC is based on the simulation of Hamiltonian dynamics as a method to generate a sequence of samples $\left\{ {{{\boldsymbol{\zeta }}^{\left( s \right)}}} \right\}_{s = 1}^S$  from a desired  $D$-variate distribution $p\left( {\boldsymbol{\zeta }} \right)$ by exploring its sample space.
%HMC is based on the simulation of Hamiltonian dynamics as a method to probe the sample space of a distribution.
 It combines gradient information of $p\left( {\boldsymbol{\zeta }} \right)$ and auxiliary variables, ${\bf p} \in R^{D \times 1} $, with density $p\left( {\bf{p}} \right) = {\cal N}\left( {{\bf{0}},{\bf{G}}} \right)$.   The Hamiltonian function is then defined as: 
\begin{equation}\label{JointLLgHam}
H\left( {{\boldsymbol{\zeta }},{\bf{p}}} \right) =   \psi \left( {\boldsymbol{\zeta }} \right) + \frac{1}{2}\log {\left( {2\pi } \right)^D}{\bf{G}} + \frac{1}{2}{{\bf{p}}^T}{\bf{G}}{\bf{p}}
\end{equation}
where $\psi \left( \boldsymbol \zeta  \right)$ is the negative log of the unnormalized $p\left( \boldsymbol \zeta \right)$ and  ${\bf{G}}$
is usually assumed to be the identity matrix. The physical analogy of \eqref{JointLLgHam} is a system with Hamiltonian dynamics which describe the total energy of the system as the sum of  the potential energy (the first term) and the kinetic energy (the last two terms). Moreover, HMC is only applicable for differentiable and unconstrained variables. However, in \eqref{eq:BayesRule}, there are some variables, ${{\boldsymbol{ \theta }}}$,  that must be positive. To handle this issue, we exploit the exponential-transformation where instead of $ \theta _q$, we use ${\phi _q}=\log ( \theta _q)$ with ${{\phi _q}}$ serving as an unconstrained auxiliary variable. Note that to use these transformations, we also need  to compute the Jacobian determinant as a result of the change of random variables.

By defining ${\boldsymbol{\zeta }} \!\!=\!\! {\left[ {{{\boldsymbol{\lambda }}^T},{\phi _0},...{\phi _{Q+1}}} \right]^T} \!\!\in {R^{(M + Q + 2) \times 1}}$ and  $p \left( \boldsymbol \zeta  \right)$ as the posterior distribution \eqref{eq:BayesRule}, the negative log of  the unnormalized $p \left( \boldsymbol \zeta  \right)$ (after the exponential-transformation)  is given by:
\begin{IEEEeqnarray}{rCl}\label{LogPost}
&\psi \left( \boldsymbol \zeta  \right)=-\log p\left( {{\boldsymbol{\lambda }},{\boldsymbol{\theta }}|{\cal D}} \right)\!\! =\!\! \sum\limits_{m = 1}^M {\sum\limits_{n = 1}^N { - {d_{c_mn}}{\lambda _m}} } 
+ {e^{{\lambda _m}}} \nonumber\\
&+ \frac{1}{2}\log \det \left( {{\bf{\tilde K}}} \right)
+\frac{1}{2} {{\boldsymbol{\lambda }}^T}{{{\bf{\tilde K}}}^{ - 1}}{\boldsymbol{\lambda }} + \sum\limits_{q = 0}^{Q+1} { - {A_q}{\phi _q} + {B_q}{e^{{\phi _q}}}}. \IEEEeqnarraynumspace
\end{IEEEeqnarray}
Also, the gradient of \eqref{LogPost} can be easily computed by using matrix derivatives~\cite{petersen2008matrix}:
\[\begin{array}{*{20}{l}}
{\frac{{\psi \left( {\boldsymbol{\zeta }} \right)}}{{\partial {\lambda _m}}} = \sum\limits_{n = 1}^N { - {d_{c_mn}}}  + N{e^{{\lambda _m}}} \!+\!\! {{\left[ {{{{\bf{\tilde K}}}^{ - 1}}{\boldsymbol{\lambda }}} \right]}_m}}\\
{\frac{{\psi \left( {\boldsymbol{\zeta }} \right)}}{{\partial {\phi _q}}}\! =\! \frac{1}{2}tr\left( {{{{\bf{\tilde K}}}^{ - 1}}\frac{{\partial {\bf{\tilde K}}}}{{\partial {\phi _q}}}} \right)\! -\! \frac{1}{2}{{\boldsymbol{\lambda }}^T}{{{\bf{\tilde K}}}^{ - 1}}\frac{{\partial {\bf{\tilde K}}}}{{\partial {\phi _q}}}{{{\bf{\tilde K}}}^{ - 1}}{\boldsymbol{\lambda }} \!\!-\!\! {A_q} + {B_q}{e^{{\phi _q}}}}.
\end{array}\]

%%%%%%%%%%%%%%%%%%%%%%%%%%%%%%%%%%%%%%%%%%%%%%%
\begin{algorithm}

	\SetAlgoLined
	\KwIn{$\boldsymbol \zeta ^{0}$}
	\KwOut{$\left\{ {{{\boldsymbol{\zeta }}^{\left( s \right)}}} \right\}_{s = 1}^S$ }
	\tcc{draw $S$  samples from $p\left( \zeta  \right)$ }
	%Compute ${{\bf{d}}_{1:n}} = {{\bf{d}}_{1:n - 1}} + {{\bf{d}}_n}$,
		Set ${{\bf{\zeta }}^{\left( 1 \right)}} = {{\bf{\zeta }}^{0}}$\;
	\For{$s \leftarrow 1$ \KwTo $S$}{

	${\bf q}^{(1)}={\boldsymbol \zeta}^{(s)}$, ${\bf{p}}^{(1)}\sim {\cal N}\left( {{\bf{0}},{\bf{G}}} \right)$\;
	Compute $H\left( {{{\bf{q}}^{(1)}},{{\bf{p}}^{(1)}}} \right)$\;
\For{$l \leftarrow 1$ \KwTo $L$}{
	${\bf{p}} = {{\bf{p}}^{(l)}} - \varepsilon \nabla \psi \left( {{{\bf{q}}^{(l)}}} \right)$\;

	${\boldsymbol{q }}^{(l+1)} = {\boldsymbol{q }}^{(l)} + \varepsilon {{\bf{G}}^{ - 1}}{\bf{p}}$\;
	${{\bf{p}}^{(l+1)}} = {\bf{p}} - \varepsilon \nabla \psi \left( {{{\bf{q}}^{(l+1)}}} \right)$\;
}
compute $dH = H\left( {\boldsymbol{q }^{(L+1)},{\bf p}^{(L+1)}} \right) - H\left( { \boldsymbol q^{(1)} ,{\bf {p}}^{(1)}}  \right)$\;
 \eIf{$rand\left( {} \right) < {e^{-dH}}$}{
$\boldsymbol \zeta^{s+1}  = {\boldsymbol q }^{(L+1)}$\tcc*{accept }
}{
	$\boldsymbol \zeta^{s+1}   = {\boldsymbol q }^{(1)}$\tcc*{reject }
}
}
	\caption{The HMC sampling algorithm}
	\label{algHMC1}
\end{algorithm}
%%%%%%%%%%%%%%%%%%%%%%%%%%%%%%%%%%%%%%%5
 The HMC sampling is depicted in Alg.\ref{algHMC1}.  Lines $5$ to $9$ represent the discretized version of the continuous Hamiltonian equations called the leapfrog method which has $2$ parameters, the number of steps $L$ and the stepsize $\epsilon$. In lines $10-15$, the new sample proposed by the Hamiltonian dynamics simulation is evaluated. If it decreases the total system energy in \eqref{JointLLgHam} it gets accepted as a new sample, otherwise it gets rejected.

%The inference problem can also be preform in on-line manner i.e. when the requests arrive sequentially. More specifically, from \eqref{eq:ObjectiveVar} we  notice  that the all historical request observations are summarized in a sum \footnote{We also note that the sufficient statistic of Poisson data is sum of all historical observation}  which  can be updated over time as ${{\bf{d}}_{1:n}} = {{\bf{d}}_{1:n - 1}} + {{\bf{d}}_n}$ (by the assumption that the sum of  data is not memory consuming). Therefore, when  new requests arrive we can just simply compute ${{\bf{d}}_{1:n}}$ as an effective  requests and  solve the same problem in \eqref{eq:opt}.  However, in order to avoid varaible reinitializing  at each time slot and improve the convergence rate, we can use the estimated variables in the last time slot as initial values for the current time slot. For example, the initial value of HMC at time slot $n$ can can set as ${\bf{\zeta }}_n^1 = {{\bf{\zeta }}_{n - 1}} = \frac{1}{S}\sum\limits_{s = 1}^S {{\bf{\zeta }}_{^{n - 1}}^{^{\left( s \right)}}}$.

The more samples there are, the more closely the distribution of the samples matches the desired  posterior distribution. Once we collect enough samples from the HMC, any function of the posterior distribution moments can be computed. Finally, the initial HMC samples are usually discarded because they may be highly correlated or far away from the true distribution. These samples are called burn-in samples. 
 %%%%%%%%%%%%%%%%%%%%%%%%%%%%%%%%%%%%%%%%%%%%%%%%

\subsection{Low complexity inference method}
Although MCMC methods asymptotically converge to the true distribution, they may not be scalable to high dimensional problems, specifically here where the number of contents is large. Therefore, in this section, we develop a low complexity algorithm for inference based on the VB method. 

For simplicity, we assume that $\boldsymbol \theta$ is an unknown deterministic hyper-parameter and then our goal is to approximate  the posterior distribution of $\boldsymbol \lambda$ and also to find a value for  $\boldsymbol \theta$ that fits to the observation best. With this assumption, the simplified posterior distribution conditioned on $\boldsymbol \theta$ is given by:
\begin{equation}\label{eq:SimPos}
p\left( {{\boldsymbol{\lambda }}|{\cal D},{\boldsymbol{\theta }}} \right) = \frac{{p\left( {{\cal D}|{\boldsymbol{\lambda }}} \right)p\left( {{\boldsymbol{\lambda }}|{\boldsymbol{\theta }}} \right)}}{{p\left( {{\cal D}|{\boldsymbol{\theta }}} \right)}}
\end{equation}
where
$p \left( {{\cal D}|{\boldsymbol{\theta }}} \right)  =  \int {p\left( {{\cal D}|{\boldsymbol{\lambda }}} \right)p\left( {{\boldsymbol{\lambda }}|{\boldsymbol{\theta }}} \right)d{\boldsymbol{\lambda }}} 
$ is the marginal likelihood. Again, there is no closed form formula for the posterior distribution in \eqref{eq:SimPos}. 

Assuming that $\boldsymbol \theta$ is given, our  goal is to approximate $p\left( {{\boldsymbol{\lambda }}|{\cal D},{\boldsymbol{\theta }}} \right)$ using the VB technique. The main idea  is to construct an approximate distribution $q({\boldsymbol{\lambda }}|{\boldsymbol{\varphi }})$ with parameters $\boldsymbol{\varphi }$ from some tractable family and then try to make this approximation be as close as possible to the true posterior distribution ${p\left( {{\boldsymbol{\lambda }}|{\cal D},\boldsymbol{\theta }} \right)}$~\cite{bishop2006pattern}. The objective is to minimize a dissimilarity measure between  ${p\left( {{\boldsymbol{\lambda }}|{\cal D}},{\boldsymbol{\theta }} \right)}$ and $q\left( {\boldsymbol{\lambda }}|\boldsymbol{\varphi } \right)$.  One  metric  to minimize is the  Kullback Leibler (KL) divergence~\cite{bishop2006pattern}:
\begin{align}\label{eq:KLMin}
&\mathop {\min }\limits_{\boldsymbol{\varphi }} \quad {\rm{KL}}\left( {q\left( {{\boldsymbol{\lambda }}|{\boldsymbol{\varphi }}} \right)||p\left( {{\boldsymbol{\lambda }}|{\cal D},{\boldsymbol{\theta }}} \right)} \right)\nonumber\\
&s.t:  {\boldsymbol{\varphi }} \in {\cal X}
\end{align}
where the minimization is taken over the parameters of the approximate distribution and ${\cal X}$ is the parameter space.
It is not difficult to see that the objective function in \eqref{eq:KLMin} can be written as:
\begin{align}\label{eq:KLdevi}
&{\rm{KL}}\left( {q\left( {{\boldsymbol{\lambda }}|{\boldsymbol{\varphi }}} \right)||p\left( {{\boldsymbol{\lambda }}|{\cal D},{\boldsymbol{\theta }}} \right)} \right) =\nonumber\\ 
&\underbrace { - {E_{q\left( {\boldsymbol{\lambda }} \right)}}\!\left[ {\log p\left( {{\cal D}|{\boldsymbol{\lambda }}} \right)p( {{\boldsymbol{\lambda }}|{\bf{\tilde K}}} )} \right] \!\!+\!\! {E_{q\left( {\boldsymbol{\lambda }} \right)}}\!\left[ {\log q\left( {\boldsymbol{\lambda }} \right)} \right]}_{{\rm{L}}\left( {{\boldsymbol{\varphi }},{\boldsymbol{\theta }}} \right)} \!+\! {\rm log} \,{p \left( {{\cal D}|{\boldsymbol{\theta }}} \right)}.
\end{align}
where ${\rm log}\,{p\left( {{\cal D}|{\boldsymbol{\theta }}} \right)}$ is independent of the variational parameter ${\boldsymbol{\varphi }}$ and therefore the minimization problem \eqref{eq:KLMin} is equivalent to minimizing ${\rm{L}}\left( {{\boldsymbol{\varphi }},{\boldsymbol{\theta }}} \right)$. Intuitively, minimizing \eqref{eq:KLdevi} incentivizes distributions
that place high mass on configurations of the approximate distribution that explain the observations well (the first term) and also rewards distributions that are entropic, meaning that they maximize uncertainty by spreading their mass on many configurations (the second term).

Now, given the simplified posterior distribution \eqref{eq:SimPos}, the objective is to find the optimal value of $\boldsymbol \theta$. The approach is based on maximum likelihood type II~\cite{mackay1992bayesian} and~\cite{rasmussen2004gaussian} where the hyper-parameter $\boldsymbol \theta$ is specified such that it maximizes the log marginal likelihood:
\begin{equation}\label{eq:marglik}
{{\boldsymbol{\theta }}^*} = \mathop {\rm argmax}\limits_{\boldsymbol{\theta }\ge 0} \quad {\mathop{{\rm log}\, p} \nolimits} \left( {{\cal D}|{\boldsymbol{\theta }}} \right)
\end{equation}
which  is not easy to compute. However, from \eqref{eq:KLdevi} it can be seen that  since the KL divergence must be zero or positive, we have
\begin{equation}\label{eq:LowerBoundLLG}
{{\rm{log}}\, p}\left( {\cal D|{\boldsymbol{\theta }}} \right) \ge  - L\left( {{\boldsymbol{ \varphi }},{\boldsymbol{\theta }}} \right).
\end{equation}

Therefore, $ -L\left( {{\boldsymbol{\varphi }},{\boldsymbol{\theta }}} \right)$ is a lower bound for the log-marginal likelihood.
This indicates that the VB provides an optimization problem for jointly finding the posterior approximation and the best values for the hyper-parameters ${\boldsymbol{\theta }}$ which is given by:
\begin{subequations}\label{eq:opt}
	\begin{align}
&\mathop {\min }\limits_{{\boldsymbol{\varphi }},{\boldsymbol{\theta }}} {\rm{L}}\left( {{\boldsymbol{\varphi }},{\boldsymbol{\theta }}} \right)\\
&s.t:{\boldsymbol{\varphi }} \in {\cal X},{\boldsymbol{\theta }}\; \ge {\bf{0}}
	\end{align}
\end{subequations}

Yet, we haven't specified the form of the approximate distribution $q\left( {\boldsymbol{\lambda }}|{\boldsymbol{\varphi }} \right)$. We employ the mean field method in which it is assumed that the variational distribution has a factorized form. Therefore, we suppose that  $q\left( {\boldsymbol{\lambda }} |{\boldsymbol{\varphi }}\right)$ has the following form:
%%%%%%%%%%%%%%%%%%%%%%%%%%%%%%%%%%%%%%%%%%%%%%%
\begin{equation}
q\left( {{\boldsymbol{\lambda }}|{\boldsymbol{\varphi }} = {{[{{\boldsymbol{\mu }}};{{\boldsymbol{\sigma }}}]}}} \right) = \prod\limits_{m = 1}^M {{\cal N}\left( {{\lambda _m}|{\mu _m},{\sigma _m}} \right)} 
\end{equation}
%%%%%%%%%%%%%%%%%%%%%%%%%%%%%%%%%%%%%%%%%%%%%%%%%%%%%
where ${\boldsymbol{\mu }} = \left[ {{\mu _1},...,{\mu _M}} \right]^T$ and ${\boldsymbol{\sigma }} = {\left[ {{\sigma _1},...,{\sigma _M}} \right]^T}$. By this assumption, ${\rm{L}}\left( {{\boldsymbol{\varphi }},{\boldsymbol{\theta }}} \right)$ in \eqref{eq:KLdevi} can be expressed as:
%%%%%%%%%%%%%%%%%%%%%%%%%%%%%%%%%%%%%%%%%
\begin{align}\label{eq:ObjectiveVar}
{\rm{L}}\left( {{\boldsymbol{\varphi }},{\boldsymbol{\theta }}} \right) =&  - {{{\bf{\bar d}}}_N}^T{\boldsymbol{\mu }} + N\sum\limits_{m = 1}^M \left\{ {e^{{\mu _m} + \frac{1}{2}{\sigma _m}}} + \frac{1}{2}{{\left[ {{{\bf{\tilde K}}^{ - 1}}} \right]}_{mm}}\!\!\!{\sigma _m}- \right.\nonumber\\ &\left.\frac{1}{2}\log {\sigma _m} \right\}  + \frac{1}{2}\left\{ {{{\boldsymbol{\mu }}^T}{\bf{\tilde K}}{^{ - 1}}{\boldsymbol{\mu }} + \log |{\bf{\tilde K}}|} \right\}
\end{align}
where ${{{\bf{\bar d}}}_N} = \sum\limits_{n = 1}^N {{{\bf{d}}_n}} $.

 %that we notice that using the variational approximation and the Jensen inequity, a lower bound for $\log Z$ can be computed.
%\begin{equation}\label{eq:JensenIneq}
%{\mathop{\rm logp}\nolimits} \left( {{\cal D}|{\boldsymbol{\theta }}} \right) = \log \int {q\left( {\boldsymbol{\lambda }} \right)\frac{{p\left( {{\cal D}|{\boldsymbol{\lambda }}} \right)p\left( {{\boldsymbol{\lambda }}|{\boldsymbol{\theta }}} \right)}}{{q\left( {\boldsymbol{\lambda }} \right)}}d{\boldsymbol{\lambda }}}  \ge \int {q\left( {\boldsymbol{\lambda }} \right){\rm log}\frac{{p\left( {{\cal D}|{\boldsymbol{\lambda }}} \right)p\left( {{\boldsymbol{\lambda }}|{\boldsymbol{\theta }}} \right)}}{{q\left( {\boldsymbol{\lambda }} \right)}}d{\boldsymbol{\lambda }}}
%\end{equation}

 The objective function in \eqref{eq:ObjectiveVar} is non-convex and therefore it is difficult to solve  \eqref{eq:opt}. However, it can be seen that \eqref{eq:ObjectiveVar}  is convex  w.t.r to $\boldsymbol{\varphi }$. To leverage this partial convexity, we apply the block-coordinate descent method~\cite[Ch.~1]{bertsekas1999nonlinear}. Iteratively, we  minimize  over block-coordinate  $\boldsymbol{\varphi }$ while fixing  $\boldsymbol \theta$  and then minimize over block-coordinate  $\boldsymbol \theta$ while fixing $\boldsymbol{\varphi }$.
 The pseudo-code of the variational block-coordinate descent method is depicted in Alg. \ref{Alg:CoordinateMethod}. %In addition, similar to Alg. \ref{algHMC1}, it can implemented for on-line data. ${{\boldsymbol{\mu }}_{n-1}},{{\boldsymbol{\sigma }}_{n - 1}},{{\boldsymbol{\theta }}_{n - 1}}$ are the parameter values of  time slot ${n - 1}$ which they can be used as initial value for  time slot $n$.
 %%%%%%%%%%%%%%%%%%%%%%%%%%%%%%%%%%%%%%%%%%%%%%%%%%
  \begin{algorithm}
  	\SetAlgoLined
  	\KwIn{${{\boldsymbol{\varphi }}^{0}},{{\boldsymbol{\theta }}^{0}}$}
  	%\KwOut{ ${{\boldsymbol{\mu }}_{n}},{{\boldsymbol{\sigma }}_{n}},{{\boldsymbol{\theta }}_{n}}$ }
  	
  	%Compute  ${{\bf{d}}_{1:n}} = {{\bf{d}}_{1:n - 1}} + {{\bf{d}}_n}$\;
  	%${{\boldsymbol{\mu }}^0} = {{\boldsymbol{\mu }}_{n - 1}},{{\boldsymbol{\sigma }}^0} = {{\boldsymbol{\sigma }}_{n - 1}},{{\boldsymbol{\theta }}^0} = {{\boldsymbol{\theta }}_{n - 1}}$\;
  	
  	\For{$t \leftarrow 1$ \KwTo convergence}{
  		$ {{{\boldsymbol{\varphi }}^t}}  = \arg \mathop {\min }\limits_{{\boldsymbol{\varphi }}} L\left( {{\boldsymbol{\varphi }},{{\boldsymbol{\theta }}^{t - 1}}} \right)$\;
  		
  		${{\boldsymbol{\theta }}^t} = \arg \mathop {\min }\limits_{{\boldsymbol{\theta }} > 0} L\left( {{{\boldsymbol{\varphi }}^t},{\boldsymbol{\theta }}} \right)$\;
  	}
  	%${{\boldsymbol{\mu }}_n} = {{\boldsymbol{\mu }}^t},{{\boldsymbol{\sigma }}_n} = {{\boldsymbol{\sigma }}^t},{{\boldsymbol{\theta }}_n} = {{\boldsymbol{\theta }}^t}$\;
  	\caption{The Variational block-coordinate descent algorithm}
  	\label{Alg:CoordinateMethod}
  \end{algorithm}
%%%%%%%%%%%%%%%%%%%%%%%%%%%%%%%%%%%%%%%%%%%%%%%%%%%

Despite Alg. \ref{Alg:CoordinateMethod}, solving  problem \eqref{eq:opt} can be challenging. More specifically, the optimization subproblem w.r.t. ${{\boldsymbol{\varphi }}}$ may be convex, but additionally it is high dimensional. As far as the subproblem w.r.t. $\boldsymbol \theta$ is concerned, it may be low dimensional, because the number of content features  is usually small, but it is non-convex. In the following, we explain how these subproblems can be efficiently solved by choosing the appropriate numerical methods.
 \begin{itemize}
 	\item Optimization w.r.t. ${{\boldsymbol{\varphi }}}$: To  efficiently solve this high dimensional convex subproblem, we choose one of the most recent parallelization techniques, the Successive Pseudo-Convex Approximation (SPCA) method~\cite{yang2017unified}, which has been shown to have a fast convergence rate. The SPCA method solves a generally non convex problem through a sequence of successively updated approximate problems to obtain a stationary point of the original one. At iteration $i$ of the algorithm, the approximation function $\widetilde {\rm{L}}\left( {{\boldsymbol{\varphi }};{{\boldsymbol{\varphi }}^{i-1}}} \right)$\footnote{For notational simplicity, we dropped $\boldsymbol \theta$ since it is fixed in this subproblem.} should satisfy the following conditions:
\begin{enumerate}[label=(\subscript{C}{{\arabic*}})]
	\item The approximate function $\widetilde {\rm{L}}\left( {{\boldsymbol{\varphi }};{{\boldsymbol{\varphi }}^{i-1}}} \right)$ is pseudo-convex in ${\boldsymbol{\varphi }}$ for any given ${\boldsymbol{\varphi }^{i-1}} \in {\cal X}$.
	\item The approximate function is continuously differentiable in ${\boldsymbol{\varphi }} \in {\cal X}$ for any given ${\boldsymbol{\varphi }^{i-1}} \in {\cal X}$ and vice versa.
	\item The gradients of $\widetilde {\rm{L}}\left( {{\boldsymbol{\varphi }};{{\boldsymbol{\varphi }}^{i - 1}}} \right)$ and the original function ${\rm{L}}\left( {\boldsymbol{\varphi }} \right)$ at ${\boldsymbol{\varphi }} = {{\boldsymbol{\varphi }}^{i-1}}$ are the same.
\end{enumerate}
Having such approximate functions, the optimization procedure is performed as  in Alg. \ref{Alg:SPCA}. The stepsize $s$ can be found by different methods, but specifically here we use the Armjio rule, a line-search technique, due to its simplicity~\cite{bertsekas1999nonlinear}. For fixed values $\gamma$ and $\eta $, with $0<\gamma,\eta  < 1$ we set $s = {\gamma ^m}$ where $m$ is the smallest non-negative integer for which:
\begin{align}
{\rm{L}}\left( {{{\boldsymbol{\varphi }}^{i - 1}}\!\! +\! {\gamma ^m}\!\left( {{{{\boldsymbol{\bar \varphi }}}^i}\! - \!{{\boldsymbol{\varphi }}^{i - 1}}} \right)} \right) &\le\! {\rm{L}}\left( {{{\boldsymbol{\varphi }}^{i - 1}}} \right) \!+ \nonumber\\
&\eta {\gamma ^m}\nabla {\rm{L}}{\left( {{{\boldsymbol{\varphi }}^{i - 1}}} \right)^T}\!\!\left( {{{{\boldsymbol{\bar \varphi }}}^i}\! -\! {{\boldsymbol{\varphi }}^{i - 1}}} \right)
\end{align}
where ${{\boldsymbol{\bar \varphi }}^i}$ is a minimizer of the approximate function ${\rm{\tilde L}}\left( {{\boldsymbol{\varphi }};{{\boldsymbol{\varphi }}^{i - 1}}} \right)$.
%%%%%%%%%%%%%%%%%%%%%%%%%%%%%%%%
\begin{algorithm}
	\SetAlgoLined
	\KwIn{${{\boldsymbol{\varphi }}^{t-1}}$}
	\KwOut{ ${{\boldsymbol{\varphi }}^{t}}$ }
	
	${{\boldsymbol{\varphi }}^0} = {{\boldsymbol{\varphi }}^{t-1}}$\;
	\For{$i \leftarrow 1$ \KwTo convergence}{
		Compute ${{\boldsymbol{\bar \varphi }}^i} = \arg \mathop {\min }\limits_{\boldsymbol{\varphi }} {\rm{\tilde L}}\left( {{\boldsymbol{\varphi }};{{\boldsymbol{\varphi }}^{i - 1}}} \right)$\;
		
	Find stepsize ${s} \in \left( {0,1} \right]$ such that ${\rm L}\left( {{{\boldsymbol{\varphi }}^{i - 1}} + {s}\left( {{{{\boldsymbol{\bar \varphi }}}^i} - {{\boldsymbol{\varphi }}^{i - 1}}} \right)} \right)$ is sufficiently small\;
	 Update ${{\boldsymbol{\varphi }}^i} = {{\boldsymbol{\varphi }}^{i - 1}} + {s}\left( {{{{\boldsymbol{\bar \varphi }}}^i} - {{\boldsymbol{\varphi }}^{i - 1}}} \right)$
	}
		${{\boldsymbol{\varphi }}^t} = {{\boldsymbol{\varphi }}^{i }}$\;
	\caption{The SPCA algorithm}
	\label{Alg:SPCA}
\end{algorithm}
%%%%%%%%%%%%%%%%%%%%%%%%%%%%%%%%%%%%%

In our setup, we consider the following approximation function  which can be easily verified that it satisfies conditions $(C1-C3)$:
\begin{align}\label{eq:psapprox}
\tilde L\left( {{\boldsymbol{\mu }},{\boldsymbol{\sigma }};{{\boldsymbol{\mu }}^{i - 1}},{{\boldsymbol{\sigma }}^{i - 1}}} \right) &= \sum\limits_{m = 1}^M \left( \tilde L\left( {{\mu _m};{\boldsymbol{\mu }}_{ - m}^{i - 1},{{\boldsymbol{\sigma }}^{i - 1}}} \right) + \right.\nonumber\\ &\left. \tilde L\left( {{\sigma _m};{{\boldsymbol{\mu }}^{i - 1}},{\boldsymbol{\sigma }}_{ - m}^{i - 1}} \right) \right) 
\end{align}
 	 	where
\begin{align}
\tilde L\!\left( {{\mu _m};{\boldsymbol{\mu }}_{ - m}^{i - 1},{{\boldsymbol{\sigma }}^{i - 1}}} \right) &\!=\!  - {\left[ {{{\bf{\bar d}}}_N} \right]_m}{\mu _m} \!+\! Ne^{\mu _m} +\nonumber\\& \frac{1}{2}{{ {{\sigma }} }^{i-1}_{m}} \!+\! \frac{1}{2}{\left[ {{\bf{\tilde K}}{{}^{ - 1}}}\!\! \right]_{mm}}\!\!\!\!\mu _m^2 \!+\! {\boldsymbol{\mu }}_{ - m}^{i - 1}{{\bf{c}}_m}{\mu _m}\nonumber\\
\tilde L\left( {{\sigma _m};{{\boldsymbol{\mu }}^{i - 1}},{\boldsymbol{\sigma }}_{ - m}^{i - 1}} \right) &\!=\! N{e^{\mu _m^{i - 1} \!+\! \frac{1}{2}{\sigma _m}}} +\nonumber\\
& \frac{1}{2}\left( {{{\left[ {{{\bf{\tilde K}}^{ - 1}}} \right]}_{mm}}\!\!\!\!{\sigma _m} \!-\! \log \left( {{\sigma _m}} \right)} \right)\nonumber
\end{align}
 	and ${{\bf{c}}_m} = {\bf{K}}_{1:m,m + 1:M}^{ - 1}$. 
 	
 	The minimization problem of \eqref{eq:psapprox} is separable and convex which can be solved  in parallel for all variables. In other words the following problems can be solved independently:
 	% first column

%\begin{minipage}{0.5\textwidth}
\begin{equation}\label{eq:sub_mu}
\mathop {\min }\limits_{{\mu _m} \in R}  \tilde L\left( {{\mu _m};{\boldsymbol{\mu }}_{ - m}^{i - 1},{{\boldsymbol{\sigma }}^{i - 1}}} \right),\quad \forall m=1,...,M
\end{equation}
%\end{minipage}% % no whitespace between the "minipage" environments
%\begin{minipage}{0.5\textwidth}
	\begin{equation}\label{eq:sub_sig}
\mathop {\min }\limits_{{\sigma _m} > 0}   \tilde L\left( {{\sigma _m};{{\boldsymbol{\mu }}^{i - 1}},{\boldsymbol{\sigma }}_{ - m}^{i - 1}} \right),\quad \forall m=1,...,M.
\end{equation}
%\end{minipage}
\\
In our implementation, we used  the  Newton method for the unconstrained problem \eqref{eq:sub_mu} and the projected Newton method for the constrained problem \eqref{eq:sub_sig} where for each problem the first and the second order derivatives can be easily computed.

 	\item Optimization w.r.t $\boldsymbol{\theta }$: To solve this non-convex problem we first transform the constrained problem, since the kernel function parameters must be positive, by the exponential-transformation ${\boldsymbol{\varphi }} = {e^{\boldsymbol{\theta }}}$. Then by using the Newton method we solve the unconstrained problem w.r.t ${\boldsymbol{\varphi }}$. However, computing the second order derivatives has two important issues. First, it is computationally expensive due to matrix inversion and multiplication of the high dimensional covariance matrix. The second one is that because of the non-convexity of the problem, the Hessian matrix may not even be positive definite. 
 	%\begin{equation}
 %{{\boldsymbol{\bar \theta }}^i} = \arg \mathop {\min }\limits_{\boldsymbol{\theta>0 }} \nabla {L_{\boldsymbol{\theta }}}\left( {{{\boldsymbol{\theta }}^{i - 1}}} \right){\boldsymbol{\theta }} + \frac{1}{{2}}{\left( {{\boldsymbol{\theta }} - {{\boldsymbol{\theta }}^{i - 1}}} \right)^T}{\nabla ^2}L\left( {\boldsymbol{\theta }} \right)\left( {{\boldsymbol{\theta }} - {{\boldsymbol{\theta }}^{i - 1}}} \right)
 %	\end{equation}
 
 %	The first order derivate is given by:
 %	\[\frac{{\partial L}}{{\partial {\theta _q}}} =  - \frac{1}{2}{{\boldsymbol{\mu }}^T}{{\bf{K}}^{ - 1}}\frac{{\partial {\bf{K}}}}{{\partial {\theta _q}}}{{\bf{K}}^{ - 1}}{\boldsymbol{\mu }} - \frac{1}{2}tr\left( {{{\bf{K}}^{ - 1}}\frac{{\partial {\bf{K}}}}{{\partial {\theta _q}}}{{\bf{K}}^{ - 1}}{\bf{\Sigma }}} \right) + tr\left( {{{\bf{K}}^{ - 1}}\frac{{\partial {\bf{K}}}}{{\partial {\theta _q}}}} \right)\]
 	
 %	The update rule is:
 %	\begin{equation}
 %{{\boldsymbol{\theta }}^{i}} = {{\boldsymbol{\theta }}^{i-1}} + \alpha \left( {{{{\boldsymbol{\bar \theta }}}^i} - {{\boldsymbol{\theta }}^{i-1}}} \right)
 %	\end{equation}

 	To mitigate the issues, we can use an approximation of the Hessian matrix. A widely utilized and successful approximation is the Broyden-Fletcher-Goldfarb-Shanno (BFGS) one. It iteratively approximates the Hessian matrix by interpolating the gradient information. The BFGS update for the $i$-th Newton iteration is:
 	\begin{equation}
{{\bf{H}}^i} = {{\bf{H}}^{i - 1}} + \frac{{{\bf{y}}{{\bf{y}}^T}}}{{{{\bf{y}}^T}{\bf{s}}}} + \frac{{{{\bf{H}}^{i - 1}}{\bf{s}}{{\bf{s}}^T}{{\bf{H}}^{i - 1}}}}{{{{\bf{s}}^T}{{\bf{H}}^{i - 1}}{\bf{s}}}}
\end{equation}
 	where ${\bf{y}} = \nabla {L_{\boldsymbol{\varphi }}}\left( {{{\boldsymbol{\varphi }}^i}} \right) - \nabla {L_{\boldsymbol{\varphi }}}\left( {{{\boldsymbol{\varphi }}^{i - 1}}} \right)$ and ${\bf{s}} = {{\boldsymbol{\varphi }}^i} - {{\boldsymbol{\varphi }}^{i - 1}}$. 
 	      In addition, it is shown that if  ${{{\bf{y}}^T}{\bf{s}}}>0$, then ${\bf H}^{i}$ is positive definite given that $\bf H^{i-1}$ is positive definite~\cite[~Ch. 8]{nocedal2006nonlinear}. This condition is called the curvature condition and it is satisfied if:
 	      \begin{equation}\label{eq:golgCon}
 	     {{\bf{y}}^T}{\bf{s}} > \left( {c{  _2} - 1} \right)\nabla {L_{\boldsymbol{\varphi }}}{\left( {{{\boldsymbol{\varphi }}^i}} \right)^T}{\bf{p}}
 	     \end{equation}
 where ${c_2} < 1$ and ${\bf{p}} =  - {{\bf{H}}^{ - 1}}{\nabla _{\boldsymbol{\varphi }}}L$.
  The Armjio rule with   condition \eqref{eq:golgCon} is used to find a stepsize for the Newton method that guarantees the positive definiteness of  ${\bf H}^i$.
  
 % {\color{red} I should mention the condition for existing the stepsize. In case the reviewers ask any bullshit.}
 \end{itemize}
%%%%%%%%%%%%%%%%%%%%%%%%%%%%%%%%%%%%%%%%%%%%%%%%%%%%%%%%%%%%%%%%%%%%%%%
%%%%%%%%%%%%%%%%%%%%%%%%%%%%%%%%%%%%%%%%%%%%%%%%%%%%%%%%%%%%%%%%%%%%%%%
\subsection{Prediction}\label{sec:prediction}
In this subsection, we explain how to  predict  future content requests. 
We define two predictive distributions for the prediction task. First, we are interested to predict the requests for the already seen contents, a task which we call \textit{Type 1 Prediction}. This can be performed using the posterior predictive distribution:
\begin{equation}\label{pospred1}
p\left( {{\bf{d}}_{c,N+1}|{\cal D}} \right) = \int {p\left( {{\bf{d}}_{c,N+1}|{\boldsymbol{\lambda }}} \right)p\left( {{\boldsymbol{\lambda }}|{\cal D}} \right)d{\boldsymbol{\lambda }}} 
\end{equation}
where ${p( {{\bf{d}}_{c,N+1}|{\boldsymbol{\lambda }}} )}$ is a Poisson distribution, the assumed generative distribution for the content requests, and ${p\left( {{\boldsymbol{\lambda }}|{\cal D}} \right)}$ is the marginal posterior distribution of ${\boldsymbol{\lambda }}$. 

Secondly, we want to predict the request for an unseen content recently introduced in the  library by the content provider. This can be computed by a second type of posterior predictive distribution, called the \textit{Type 2 Prediction}, defined as:
\begin{align}\label{eq:pospredtype2}
&p\left( {{d_{c_{new},N+1}}|{{\bf{x}}_{new}},{\cal D}} \right) \!= \nonumber \\
 &\!\!\int \!\!\!p\left( {{d_{c_{new},N+1}}|\!{\lambda _{new}}} \right)
 \!p\left( {{\lambda _{new}}|{\boldsymbol{\lambda }},{\boldsymbol{\theta }},\!{{\bf{x}}_{new}}} \right)\! 
 p\left( {{\boldsymbol{\lambda }},\!{\boldsymbol{\theta }}|\!{\cal D}} \right)\!d{\lambda _{new}}d{\boldsymbol{\lambda }}d{\boldsymbol{\theta }}
\end{align}
where ${{\bf{x}}_{new}}$ is the feature vector of the unseen content and ${p\left( {{\lambda _{new}}|{\boldsymbol{\lambda }},{\boldsymbol{\theta }},{{\bf{x}}_{new}}} \right)}$ is a Normal distribution\footnote{Note that the generative model for the content popularities is a Gaussian process and based on its properties any finite conditional distribution is a Normal distribution.}  with mean and variance:
 \begin{align}
&{{\hat \lambda }_{new}}\! =\! {{{\bf{\tilde k}}}^T}{{{\bf{\tilde K}}}^{ - 1}}{\boldsymbol{\lambda }} \\
&{{\hat \sigma }_{new}} = K\left( {{{\bf{x}}_{new}},{{\bf{x}}_{new}}} \right) + {\theta _0} - {{{\bf{\tilde k}}}^T}{{{\bf{\tilde K}}}^{ - 1}}{\bf{\tilde k}}.
\end{align}
where  ${\bf{\tilde k}} = \left[ {K\left( {{{\bf{x}}_1},{{\bf{x}}_{new}}} \right),...,K\left( {{{\bf{x}}_M},{{\bf{x}}_{new}}} \right)} \right]{^T}$.
%Since  ${p\left( {{\boldsymbol{\lambda }}|{\cal D}} \right)}$ is not in a closed-form,  we don't have analytical expression for \eqref{pospred1}. But, it can be  approximated with a discrete distribution by HMC samples.

However, we wish to make point predictions rather than dealing with the whole predictive distribution. The best guess for a point estimate in the Bayesian context is based on risk (or loss) minimization~\cite[Chapter~2]{robert2007bayesian}. In other words, a loss function is defined which specifies the loss incurred by guessing values ${\bf{d}}_{c,N+1}$ and ${{d_{c_{new},N + 1}}}$ when the actual values are ${\bf{d}}_{c,N + 1}^*$ and $d_{c_{new},N + 1}^*$. The most common loss evaluation metric is the quadratic loss. The values of ${\bf{d}}_{c,N+1}$ and ${{d_{c_{new},N + 1}}}$ that minimize this risk function are the means of the predictive distributions and they are respectively approximated by the HMC and the VB methods as:

%: It should be noted that \eqref{eq:pospredtype2} is the distribution of the natural parameter of a new content. The point estimation of the request rate can be approximated as:

\begin{itemize}
	\item HMC:
	\begin{equation}\label{mean_app}
	E\left\{ { {{\bf{d}}_{c,N + 1}}|{\cal D}} \right\} \approx  \frac{1}{S}\sum\limits_{s = 1}^S {{e^{{{\boldsymbol{\lambda }}^{(s)}}}}}
	\end{equation}
	\begin{equation}\label{eq:mean_apptyep2}
E\left( {{d_{c_{new},N + 1}}|{{\bf{x}}_{new}}},{\cal D} \right) \approx \frac{1}{S}\sum\limits_{s = 1}^S {{e^{\hat \lambda _{new}^{\left( s \right)} + \frac{1}{2}\hat \sigma _{new}^{\left( s \right)}}}}
	\end{equation}
	\item VB:
	\begin{equation}
	E\left\{ { {{\bf{d}}_{c,N + 1}}|{\cal D}} \right\}\approx {e^{{\boldsymbol{\mu }} + \frac{1}{2}{\boldsymbol{\sigma }}}}
	\end{equation}
	\begin{equation}
	E\left( {{d_{c_{new},N + 1}}|{{\bf{x}}_{new}}},{\cal D} \right)\approx {e^{{\bar \lambda _{new}} + \frac{1}{2}{{\bar \sigma }_{new}}}}
	\end{equation}
	where
	%%%%%%%%%%%%%%%%%%%%%%%%%%%%%%
\[\begin{array}{*{20}{l}}
{{{\bar \lambda }_{new}} = {{{\bf{\tilde k}}}^T}{{{\bf{\tilde K}}}^{ - 1}}{\boldsymbol{\mu }}}\\
{{{\bar \sigma }_{new}} = {{\hat \sigma }_{new}} - {{{\bf{\tilde k}}}^T}{{{\bf{\tilde K}}}^{ - 1}}{\bf{\Sigma }}{{{\bf{\tilde K}}}^{ - 1}}{\bf{\tilde k}},\quad}\\
{\bf{\Sigma }} = {\rm{Diag}}\left( {{\sigma _1},....,{\sigma _M}} \right)
\end{array}\]	
\end{itemize}
These posterior predictive mean estimates based on HMC and VB are basically the output of the popularity prediction process which are subsequently utilized in the cache placement policy as described in Fig \ref{fig:flowchartLearningCaching}. 
 
\section{Simulation Results}\label{SimRes}
In this section, we present our simulation results to show the performance of the proposed probabilistic content request model denoted by "\textit{PGP}". To compare our results, we use as benchmark methods the ones suggested in~\cite{doan2018content} and~\cite{bharath2016learning}. Specifically, the authors in~\cite{doan2018content} used several regression methods in order to predict the popularity of  contents based on their similarity with the seen contents. In the simulations, we use the support vector regression (SVR) with  the radial basis kernel which they showed that it provides the best prediction performance for  unseen contents. In addition, in~\cite{bharath2016learning}, the classical approach of Independent Poisson MLE popularity learning is presented.

As far as the Monte Carlo  simulations are concerned, for each parameter setting scenario, where the parameters are explained later on, we run $50$ simulations. Also, for the HMC technique, we set $\varepsilon =.015$ and $L=20$ and run it for 5000 samples where the first 2500 samples were considered as the burn-in samples. Moreover, in all our simulations, we assume that the total number of contents is split in two parts, $M$ for the seen contents and $25\%$ of $M$ for the unseen contents. The simulation results are divided in two subsections: $i)$ the popularity prediction accuracy and $ii)$ the CHR gain performance. Additionally,  $2$ types of data are used for our simulations, synthetically generated and real-world data. In the first one, we assume that we know the request generation process and we therefore can synthetically generate them. This allows us to evaluate the accuracy of our popularity estimates, since we know beforehand the true model parameter values. In the second one, real-world observations are used where our model is applied on, but since we do not know the true model parameters, their evaluation is not possible. Thus, real-world content request data will only be considered in the CHR gain performance subsection.
\subsection{Popularity prediction performance}\label{subseq:preformance}

In this subsection, we compare in terms of popularity prediction accuracy our model with~\cite{doan2018content} as a benchmark using only synthetic data. To synthetically generate content requests, we use model \eqref{eq:CellGPdef}. The number of features is $Q=4$ and specifically features ${x}_m^{(1)}$, ${x}_m^{(2)}$ and ${x}_m^{(3)}$ are binary whose values are randomly generated from Bernoulli distributions with parameters $0.5$, $0.8$ and $0.2$ for all $m$ respectively. Feature $x_m^{(4)}$ is continuous and generated from a Normal distribution with zero mean and unit variance for all $m$. Moreover, we set ${\eta} = .0001,{\alpha _0} = 0.1,{\alpha _1} = 0.25,{\alpha _2} = 0,{\alpha _3} = 0.1$ and ${\alpha _4} = 0.5$.

\begin{figure}
 	\centering
 	\includegraphics[scale=0.48, trim=24 0 0 35]{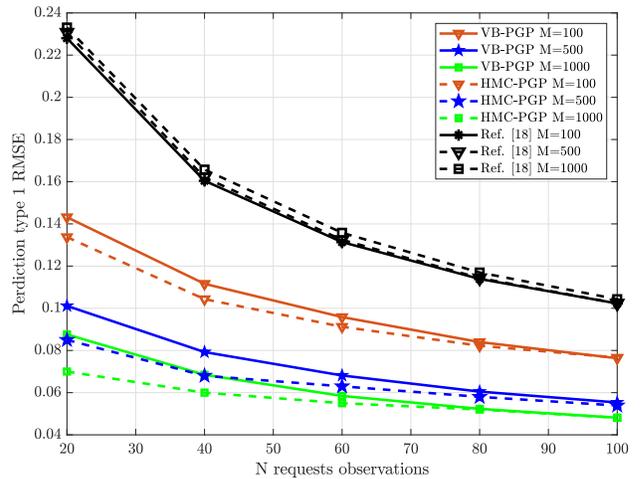}
 	\caption{Prediction type 1  RMSE vs $N$ request observations}
 	\label{fig:RMSEtype1}
 \end{figure}

Fig. \ref{fig:RMSEtype1} shows the root mean square error (RMSE) of the Type 1 popularity prediction for seen contents defined by \eqref{pospred1} vs the number of content request observations, $N$, for different content numbers, $M$. It is shown that the Bayesian PGP model predicts the requests of already seen content  significantly better than the method in~\cite{doan2018content}. Here, we note that the method in~\cite{doan2018content} is equivalent to the MLE approach  for seen contents and the content features are only used to predict the popularity of unseen contents. 
In addition, the HMC based inference performs slightly better than the VB based one. We can also observe that as $M$ increases, the accuracy of the PGP model improves. This is because the Gaussian process can  learn better the relationship between the popularities and the features.
 %%%%%%%%%%%%%%%%%%%%%%%%%%%%%%%%%%%%
%%%%%%%%%%%%%%%%%%%%%%%%%%%%%%%%%%%%%%%

\begin{figure}
	\centering
	\includegraphics[scale=0.48, trim=24 0 0 35]{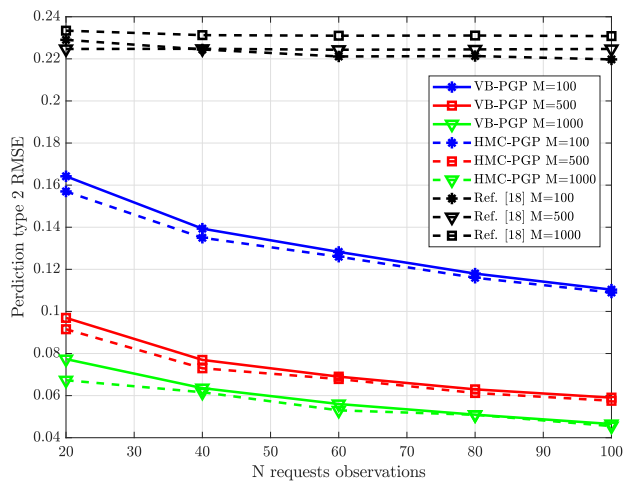}
	\caption{Prediction type 2 RMSE vs $N$ request observations}
	\label{fig:RMSEtype2}
\end{figure}

Fig. \ref{fig:RMSEtype2} shows the RMSE of the predicted popularity of unseen contents, Type 2 popularity prediction \eqref{eq:pospredtype2}, vs the number of observations, $N$. The features of the unseen contents are randomly generated with the same process as for the existing ones. As we see, the performance of our prediction algorithm is considerably better than the one in~\cite{doan2018content}. Again, similar to Fig. \ref{fig:RMSEtype1}, the prediction accuracy improves when either $N$ or $M$ increases and also the HMC based PGP is a bit more accurate than the VB based PGP.
%%%%%%%%%%%%%%%%%%%%%%%%%%%

Next, we evaluate the performance of the HMC and the VB inference methods in Tables \ref{table:Tab1} and \ref{table:Tab2}. Here, we show the estimated mean values of the kernel function parameters. As we expected, it is observed that as the number of observations increases we get closer to the true values. However, from the tables, the estimation accuracy improvement of the parameters is largely affected by the number of contents. For example, for feature $x_m^{(2)}$, which does not affect the popularities, the estimation of its scale variation, ${\alpha _2}$, is better at $N=80$ for $M=500$ in comparison with $M=100$. Moreover, it can be seen that the HMC has better estimation accuracy with respect to the VB. These results confirm our previous simulations where as $M$ increases, the Gaussian process gets more accurate and consequently shows a better prediction performance.

% In this figure and the previous ones, we observe that there is a trade-off between the HMC and the VB performances. In one hand, the HMC is more accurate but it converges slowly. On the other hand, the VB is less accurate but faster than the HMC.

 \begin{table} 
	\centering
	\begin{tabular}{|c|c|c|c|c|c|c|}
		\hline
		\cline{2-5}
		&\multicolumn{2}{|c|}{N=20}&\multicolumn{2}{|c|}{N=80}& \\
		\cline{2-5}
		& HMC & VB & HMC & VB &True value \\
		\hline
		\multicolumn{1}{|c|}{${\eta }$} & 0.0044 & 0.0232 & 0.0021 & 0.0073 &0.0001  \\
		\hline
			\multicolumn{1}{|c|}{${\alpha _0}$} &0.1299 & 0.0888 &0.1258  & 0.1519 & 0.1 \\
		\hline
			\multicolumn{1}{|c|}{${\alpha _1}$} & 0.2179& 0.1819 &0.2259   & 0.2289 & 0.25 \\
		\hline
			\multicolumn{1}{|c|}{${\alpha _2}$} &0.0199 & 0.0502 &  0.0109&0.0086&   0 \\
		\hline
		\multicolumn{1}{|c|}{${\alpha _3}$} &0.0961 & 766.6762 &0.0642  & 0.0591&  0.1  \\
						\hline
		\multicolumn{1}{|c|}{${\alpha _4}$} &0.3948 & 0.4009 & 0.4375 & 0.4266 & 0.5 \\
		\hline
	\end{tabular}
\caption{The estimates of the kernel function parameters for $M =100$}
	\label{table:Tab1}
\end{table}
\begin{table}
	\centering
	\begin{tabular}{|c|c|c|c|c|c|c|}
		\hline
		\cline{2-5}
		&\multicolumn{2}{|c|}{N=20}&\multicolumn{2}{|c|}{N=80}&\\
		\cline{2-5}
		& HMC & VB & HMC & VB& True value \\
		\hline
		\multicolumn{1}{|c|}{${\eta}$} & 0.0013  & 0.0160 & 0.0002 & 0.0052& 0.0001  \\
		\hline
		\multicolumn{1}{|c|}{${\alpha _0}$} &0.1260 & 0.1295 &0.1221  & 0.1388&0.1   \\
		\hline
		\multicolumn{1}{|c|}{${\alpha _1}$} & 0.2382& 0.1568 &0.2298   & 0.1963 & 0.25 \\
		\hline
		\multicolumn{1}{|c|}{${\alpha _2}$} &0.0035 & 0.0041  &  0.0031& 0.0032&  0 \\
		\hline
		\multicolumn{1}{|c|}{${\alpha _3}$} &0.0751 & 0.0626 &0.0936  & 0.0799&  0.1 \\
		\hline
		\multicolumn{1}{|c|}{${\alpha _4}$} &0.4890 & 0.4069  & 0.4986 & 0.4274&  0.5 \\
		\hline
	\end{tabular}
	\caption{The estimates of the kernel function parameters for $M =500$}
	\label{table:Tab2}
\end{table}
%%%%%%%%%%%%%%%%%%%%%%%%%%%%%%%%%%%%%%%%%%
%%%%%%%%%%%%%%%%%%%%%%%%%%%%%%%%%%%%%%%%%
%%%%%%%%%%%%%%%%%%%%%%%%%%%%%%%%%%%%%%%%%%%
\subsection{Caching gain Performance}\label{subseq:cachegain}
In this subsection, we investigate how the prediction performance of our model affects the CHR when using the caching policy defined in \eqref{eq:cachePolicy}. Throughout this subsection, we  set $M=500$ and $N=40$  unless otherwise stated. In addition, the cache capacity is shown as the percentage of the total size of all contents.
 In the following, we respectively evaluate the CHR on synthetically generated requests and a real-world dataset.
 
\begin{figure}[htbp]
	\centering
	\begin{subfigure}[b]{0.35\textwidth}
		 \centering
		\includegraphics[scale=0.48, trim=90 -15 0 40]{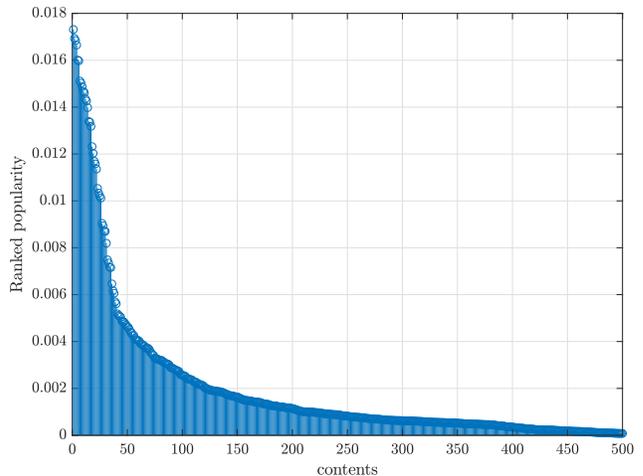}
		\caption{$\alpha_0=5$ }
		\label{fig:gull}
	\end{subfigure}

	\begin{subfigure}[b]{0.35\textwidth}
		\centering
		\includegraphics[scale=0.48, trim=90 -15 0 0]{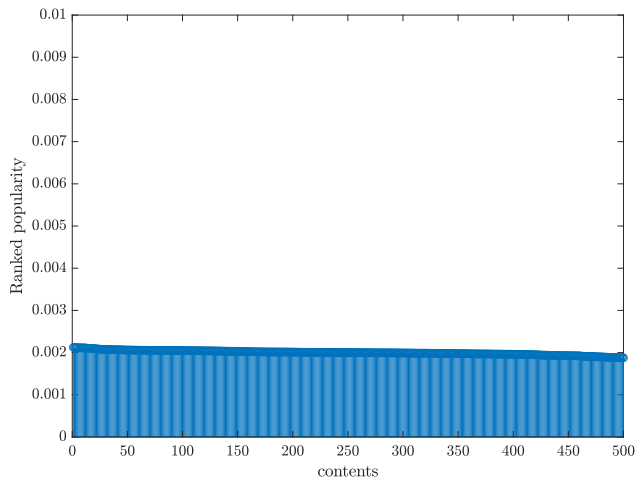}
		\caption{$\alpha_0=0.001$}
		\label{fig:tiger}
	\end{subfigure}
	\caption{Samples from a Gaussian process to generate content popularities}\label{fig:RankPop}
\end{figure}

To generate synthetic data, we use the model \eqref{eq:GPuserLevel} rather than  \eqref{eq:CellGPdef}, since it gives more flexibility to study the influence of users' behavior on the CHR. For simplicity, we assume that the parameters of the kernel function in \eqref{eq:userKernel} are the same for all users with ${\beta _{pu}} = 1$  and also the content features are generated as previously.  The $P$ user features are generated randomly as:
%%%%%%%%%%%%%%%%%%%%%%%%%%%%%%%%%%%%
\begin{equation}\label{eq:Dir}
{{\bf{p}}_u}\sim {\rm Dirichlet}\left( \omega  \right), \quad \forall u = 1,...,U
\end{equation}
%%%%%%%%%%%%%%%%%%%%%%%%%%%%%%%%%%
where ${\rm Dirichlet}\left( \omega  \right)$ is a symmetric Dirichlet distribution with parameter  $\omega$. By properly tuning $\omega$, we can control the  similarity between  user features. More specifically, if $\omega$ is set to be large, then the generated samples from \eqref{eq:Dir} are very much alike which means that all users have more or less the same features and thus become highly correlated. On the other hand, if $\omega$ is set to be small, then the generated samples  correspond to the sparse case in which each user only has a small number of features. In this case, by setting $P$ to be large, users almost have non-overlapping features which we can assume more or less they are dissimilar. In the simulations, we set $P=100$.  
Moreover the number of users, $U$, is $10$ and the sizes of the contents are randomly generated from the interval $\left( {0,100} \right)$.  In addition, we can generate popularities which mimic the Zipf distribution by tuning $\alpha_0$ while the other parameters are fixed. 
For example, Fig. \ref{fig:RankPop} depicts the ranked normalized versions of the following generated  popularities: 
\begin{equation}\label{eq:NormPop}
{r_m}\left( {{{\bf{x}}_m}} \right) = \sum\limits_{u = 1}^U {{e^{{\lambda _m}\left( {{{\bf{x}}_m},{{\bf{p}}_u}} \right)}}},\quad \forall m=1,...,M
\end{equation}
with $\omega=1$ and different values of $\alpha_0$. The figures show that as $\alpha_0$ decreases the distribution of the ranked popularities converges to a Zipf distribution with small peakiness parameter. 
%%%%%%%%%%%%%%%%%%%%%%%%%%%%%%%%%%%%%%%%%%
%%%%%%%%%%%%%%%%%%%%%%%%%%%%%%%

\begin{figure}
	\centering
	\includegraphics[scale=0.48, trim=24 0 0 40]{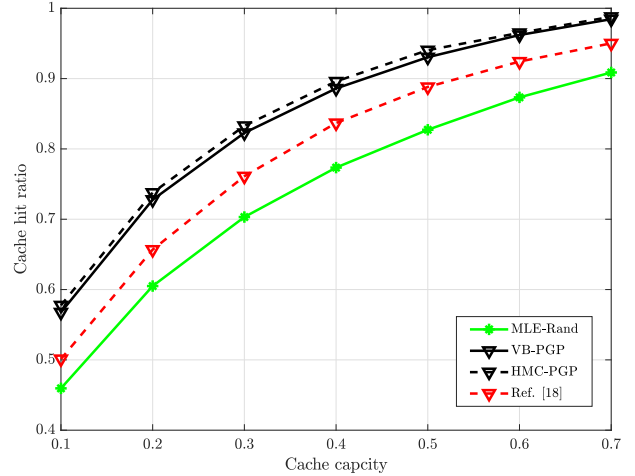}
	\caption{CHR vs cache capacity for $\alpha_0=2.5,\omega=1$}
	\label{fig:CHRVerCap}
\end{figure}

In all the subsequent figures, a partially random benchmark caching strategy denoted by "\textit{MLE-Rand}" is also depicted in which we split the cache memory in two parts: $80\%$ is assigned for the seen contents with their popularities estimated by \eqref{eq:MeanPoisson} and $20\%$ for randomly selecting the unseen contents\footnote{We chose this percentage because of the assumed percentage number  of the seen and unseen contents}. Fig. \ref{fig:CHRVerCap} shows the CHR vs the cache capacity for $\alpha_0=2.5$ and $\omega =1$. As we expected, CHR increases as the cache capacity increases. It can be observed that the PGP model based caching outperforms the other caching methods in this cache capacity range. For example, for cache capacity at $0.3$, the VB-PGP and HMC-PGP assisted caching improve the CHR by $8\%$ and $17\%$ with respect to~\cite{doan2018content} and MLE-Rand methods respectively. In addition, both CHR performances based on our models, HMC-PGP and VB-PGP, are very close to each other.

\begin{figure}
	\centering
	\includegraphics[scale=0.48, trim=24 0 0 40]{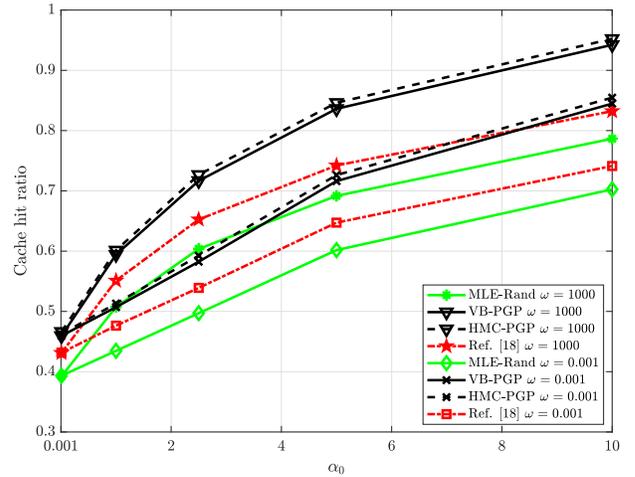}
	\caption{CHR vs $\alpha_0$}
	\label{fig:CHRVerUser}
\end{figure}

Fig. \ref{fig:CHRVerUser} illustrates CHR versus $\alpha_0$  for different values of $\omega$. The cache capacity is $0.2$  of the total size of contents. It can be seen that as $\alpha_0$ increases the CHR also gradually increases. This is expected since as $\alpha_0$ increases the distribution of content popularities becomes more like a Zipf with a large peakiness parameter (Fig. \ref{fig:RankPop}) meaning that only a few contents have significant contribution to the CHR gain and these can be easily distinguished since they are requested more than the other contents. On the other hand, for small $\alpha_0$ values the distribution of content popularities will be more like a uniform or a Zipf with a small peakiness parameter indicating that all contents have almost the same contribution to the CHR gain. Moreover, for a fixed and relatively large value of $\alpha_0$, the CHR decreases as $\omega$ decreases. This can be explained by noticing that when $\omega$ decreases the users will have different content preferences since they become uncorrelated. Therefore, by aggregating all the user requests, popularities are almost brought into uniformity indicating less CHR gain. In addition, we observe that the caching assisted by both of our methods is superior to the other ones for all  scenarios. More specifically, the performance gap between the PGP based caching and the other ones increases as $\alpha_0$ increases. This is because for large $\alpha_0$, the CHR is very sensitive to the prediction accuracy and a small prediction mistake may cause a huge CHR reduction. This is also supported by our results in the previous subsection where we showed that our model provides higher prediction accuracy with respect to the other methods. On the other hand, for small  $\alpha_0$ the prediction accuracy will have a small effect on the CHR. %The same conclusion can be drawn with regard to different values of $\omega$.

\begin{figure}
	\centering
	\includegraphics[scale=0.48, trim=24 0 0 40]{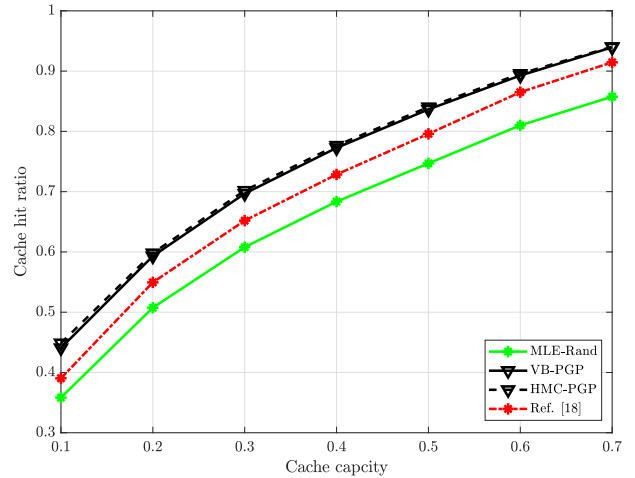}
	\caption{CHR vs cache capacity on MovieLens Dataset }
	\label{fig:CHRMovieLens}
\end{figure}

Finally, we show the CHR performance based on our model on the real-world MovieLens 20M dataset~\cite{harper2016movielens}. This consists of $18$ movie genres (action, adventure, animation, children's, comedy and so on). From the dataset, we choose ratings over 2 years, 2010 and 2011. Similar to~\cite{muller2017context}, a movie's rate is considered as one request for this movie. The length of the time slots is considered as one day. In addition, we observed that the movie popularities are almost constant during every two months of this period which indicates that our model can be applied in order to learn (almost) stationary content request distributions as underlined in the beginning of Section \ref{seq:Sysmodel}. Therefore, in our simulations, the 2-year time interval is separated in $12$ bimonthly intervals where for each one the model is trained with the request observations of 30 days ($N=30$) and the CHR is evaluated during the next 30 days. Fig. \ref{fig:CHRMovieLens} illustrates the CHR vs the cache capacity. Again, it can be seen that our model has a better prediction accuracy and consequently improved CHR with respect to both the MLE-Rand method and the one presented in~\cite{doan2018content}. For instance, for the cache capacity at $0.3$, our method improves CHR by $6\%$ and $11\%$ compared to the method in~\cite{doan2018content} and the  MLE-Rand method respectively.  %The figure clearly shows that in practice  movie features contain useful information about the underlying request which need to be leveraged in order to improve the CHR.
 
%\subsection{Discussion}
%In this paper, we mainly focused on developing a  probabilistic model for content requests and to evaluate the prediction performance of the model, we used a common caching policy. However, throughout our simulations, we found out that with this simple policy the prediction accuracy is not translated well into the CHR gain.  The reason is that the policy in \eqref{eq:cachePolicy} cares  only about a few popular contents and the prediction accuracy of the rest is ignored. In order to fully leverage the prediction accuracy a more sophisticated policy is needed e.g. considering the  wireless physical layer features. Nevertheless, still with this simple caching policy, we obtained around $5\%$ CHR gain with respect to~\cite{doan2018content} which is a significant improvement considering that the traffic load has  an accumulate effect from the edge to the core network. For example, assume that there are 5 micore BSs connecting to a macro BS through the backhaul links. Moreover, for each micro BS, we reduce $5\%$ the traffic load on its backhaul link. This will reduce $25\%$ data traffic from the macro BS to the core network which is huge.
%%%%%%%%%%%%%%%%%%%%%%%%%%%%%%%%%%%%%%%%%%%%%

\section{Conclusions}\label{Conclusion}
In this paper, we proposed a flexible model for modeling the content requests and predicting their popularity. We proposed a multilevel probabilistic model, the Poisson regressor based on a Gaussian process, that can exploit the content features and provide accurate popularity estimation. For the seen contents, the Gaussian process acts as a regularizer which results in better estimation of their popularities. When  new unseen content is introduced in the library, the proposed model can predict its popularity based on its correlation with the existing seen contents. We utilized Bayesian learning to obtain the parameters of the model because it is robust against overfitting and therefore efficient in edge-caching systems where overfitting is a big challenge due to the small number of request observations. Because the posterior distribution is not in closed form, we resort to approximation methods: the HMC sampling and the VB learning.  In the simulation results, we showed that both techniques have good performance for our PGP model. Particularly, the VB is less computationally intensive than the HMC, but also a bit less accurate. Finally, we compared our method with the state-of-the-art popularity estimation scheme and  showed that our method improves performance significantly.
\section*{Acknowledgment}
This work was funded by the National Research Fund (FNR),
Luxembourg under the projects "LISTEN" and "PROCAST". This work was also supported  by the European Research Council (ERC) under the project "AGNOSTIC".
%%%%%%%%%%%%%%%%%%%%%%%%%%%%%%%%%%%%%%%%%%%%%%%%%%
%\bibliography{mybib}{}
\bibliographystyle{IEEEtran}

\end{document}